\documentclass[useAMS,usenatbib]{mn2e}
\usepackage{xspace}
\usepackage[section]{placeins}
\usepackage{graphicx,txfonts}

\makeatletter
\g@addto@macro\@floatboxreset\centering
\makeatother 

\usepackage{hyperref}

\usepackage[usenames,dvipsnames]{color}

\newcommand{\Sref}[1]{Section~\ref{#1}}

\sfcode`\.=1001\sfcode`\?=1001\sfcode`\!=1001
\newcommand{\Fref}[1]{\ifhmode \ifnum\spacefactor=1001 Figure~\ref{#1}\else Fig.~\ref{#1}\fi \else Figure~\ref{#1}\fi}
\newcommand{\Eref}[1]{\ifhmode \ifnum\spacefactor=1001 Equation~(\ref{#1})\else equation~(\ref{#1})\fi \else Equation~(\ref{#1})\fi}

\newcommand{\ms}{\hbox{${\rm m\,s}^{-1}$}\xspace}
\newcommand{\msperthousand}{\hbox{${\rm m\,s}^{-1}{\rm \,per\,}1000$\,\AA}\xspace}
\newcommand{\kms}{\hbox{${\rm km\,s}^{-1}$}\xspace}
\newcommand{\SNR}{\hbox{${\rm SNR}$}\xspace}

\newcommand{\zab}{\ensuremath{z_\textrm{\scriptsize abs}}}

\newcommand{\ion}[2]{\ensuremath{\textrm{#1\,{\scshape{#2}}}}} 
\newcommand{\logN}{\ensuremath{\log N}\xspace}

\newcommand{\vshift}{\hbox{$v_{\textrm{\scriptsize shift}}$}\xspace}
\newcommand{\bspsmall}{\vspace{0.5cm}\small\noindent This paper has been typeset from a \TeX/\LaTeX\ file prepared by the author.\normalsize}
\newcommand{\daa}{\ensuremath{\Delta \alpha/\alpha}\xspace}

\title[Systematic errors in quasar spectra]{Impact of instrumental systematic errors on fine-structure constant measurements with quasar spectra}

\author[J.~B.~Whitmore \& M.~T.~Murphy]{Jonathan B. Whitmore,$^{1}$\thanks{E-mail:
    jbwhit@gmail.com (JBW)} Michael T. Murphy$^{1}$ \\
  $^{1}$Centre for Astrophysics and Supercomputing, Swinburne University of Technology, Hawthorn, Victoria 3122, Australia\\
 }
\begin{document}

\date{Accepted ---.  Received ---; in original form 2014 September 16}

\pagerange{\pageref{firstpage}--\pageref{lastpage}}
\pubyear{2014}
\maketitle

\label{firstpage}

\begin{abstract}
  We present a new `supercalibration' technique for measuring
  systematic distortions in the wavelength scales of high resolution
  spectrographs. By comparing spectra of `solar twin' stars or
  asteroids with a reference laboratory solar spectrum, distortions in
  the standard thorium--argon calibration can be tracked with
  $\sim$10\,m\,s$^{-1}$ precision over the entire optical wavelength
  range on scales of both echelle orders ($\sim$50--100\,\AA) and
  entire spectrographs arms ($\sim$1000--3000\,\AA). Using archival
  spectra from the past 20 years we have probed the supercalibration
  history of the VLT--UVES and Keck--HIRES spectrographs. We find that
  systematic errors in their wavelength scales are ubiquitous and
  substantial, with long-range distortions varying between typically
  $\pm$200\,m\,s$^{-1}$\,per 1000\,\AA. We apply a simple model of
  these distortions to simulated spectra that characterize the large
  UVES and HIRES quasar samples which previously indicated possible
  evidence for cosmological variations in the fine-structure constant,
  $\alpha$.  The spurious deviations in $\alpha$ produced by the model
  closely match important aspects of the VLT--UVES quasar results at
  all redshifts and partially explain the HIRES results, though not
  self-consistently at all redshifts. That is, the apparent ubiquity,
  size and general characteristics of the distortions are capable of
  significantly weakening the evidence for variations in $\alpha$ from
  quasar absorption lines.
\end{abstract}

\begin{keywords}
  atomic data -- line: profiles -- techniques: spectroscopic --
  methods: data analysis -- quasars: absorption lines
\end{keywords}

\section{Introduction} 
\label{sec:introduction}
    The Standard Model of particle physics requires a number of fundamental physical constants as inputs.
    These constants are defined by dimensionless ratios of physical quantities, e.g.\ the charge of the electron, or the speed of light, but, as dimensionless ratios, they have the same value for any choice of physical units.
    No physical theory predicts their values, so we must derive their values from measurements.
    The fine-structure constant, $\alpha \equiv e^2 / \hbar c$, is the coupling constant that sets the scale of electromagnetic interactions, and its possible variation will be the focus of this paper:
    \[
        \daa \equiv \frac{\alpha_z - \alpha_0}{\alpha_0},
    \]
    with the value of $\alpha$ at some redshift $z$\ denoted as $\alpha_z$ and the current laboratory value as $\alpha_0$.

    Experimental and observational limits have been placed on $\daa$ over a range of precisions and time-scales, with \citet{2011LRR....14....2U} reviewing the various methods and constraints.
    These generally range from a very tight constraint over several years in laboratories, e.g.~a few parts per 10$^{17}$ \citep[e.g.][]{2008Sci...319.1808R}, to somewhat looser constraints at cosmological scales, e.g.~a few percent with the cosmic microwave background \citep[e.g.][]{Menegoni:2012tf}.
    However, measurements of \daa in absorption systems found within quasar spectra probe values of $\alpha$ over cosmological time and distance with a typical precision of several parts per million (ppm).
    The interest in the possibility of the variation of two fundamental constants ($\alpha$ and $\mu$) has intensified over the past 15 years since the emergence of some evidence for a cosmological variation of $\alpha$ from quasar studies.
    This evidence came with the increased sensitivity to $\alpha$ variation enable by the `Many-Multiplet' (MM) method \citep{Dzuba:1999,Webb:1999cy}: the comparison of many different metal ion transitions whose frequencies have widely differing dependence on $\alpha$.
    Absorption systems that lie along the line-of-sight to a quasar imprint a number of narrow metal absorption lines onto its spectrum, and \daa is measured by comparing the relative wavelength spacing of these features.
    
    The MM method was first applied to a sample of 30 quasar absorption systems with the Keck--HIRES in \citet{Webb:1999cy}.
    In the years since, the MM method has been applied to several other individual absorption systems \citep[e.g.,][]{Quast:2004ip,Levshakov:2007ch,Molaro:2008jt,2013A&A...555A..68M}. 
    However, the most statistically significant results have come from two large samples: a substantially increased Keck--HIRES sample and a more recent sample measured with the VLT--UVES spectrograph.
    HIRES and UVES are high resolution ($R\approx50,000$--80,000) echelle spectrographs on 8-to-10-m class telescopes: Keck (Hawaii) and VLT (Chile) respectively.
    The final Keck sample combined $\daa$ measurements from 140 absorption systems, along 78 lines of sight, to yield a statistically significant non-zero weighted average of $\daa = -5.7 \pm 1.1$ ppm \citep{Murphy:2003em,Murphy:2004kl}.
    \citet{2012MNRAS.422.3370K} later analyzed 153 systems with the VLT and found a statistically significant positive average value. 
    However, by combining the VLT sample with the Keck sample, \citet{2011PhRvL.107s1101W} and \citet{2012MNRAS.422.3370K} found evidence for a spatial dipole in the value of $\alpha$ across the sky.
    These surprising results demand detailed investigations into possible systematic effects in high resolution spectrographs that might mimic this possible evidence for a varying $\alpha$ \citep[e.g.][]{Murphy:2001uh,Murphy:2003em,Molaro:2008hc,Griest:2009kpa,2010ApJ...723...89W,2012MNRAS.422.3370K,2013A&A...555A..68M,2013MNRAS.435..861R,2013ApJ...778..173E,Evans:2014:128}.
    We continue that effort here.
    
    Any significant distortion of the wavelength scale, especially over long wavelength ranges ($\gtrsim$1000\,\AA), could have a serious impact on the accuracy of \daa measurements from quasar absorption spectra.
    The predicted relative wavelength shifts for $\daa \sim $ few ppm correspond to 1/10th of a pixel wavelength accuracy across the wavelength range of the spectrograph.
    Thus, having an accurate wavelength scale is extremely important.
    The standard method used to calibrate the wavelength scale of quasar exposures on high resolution spectrographs is by comparison with a separate ThAr arc lamp exposure.
    The sharp ThAr emission lines have known wavelengths, and the positions at which they fall on the CCD are used to create a wavelength solution for the separate quasar exposure.

    There have been a number of tests of the wavelength solution provided by ThAr spectra, especially with regards to fine-structure constant work, including \citet{1995PASP..107..966V}, \citet{Murphy:2001uh}, \citet{Molaro:2008hc}, \citet{Griest:2009kpa}, \citet{2010ApJ...723...89W}, \citet{Wilken:2010fc}, \citet{WendtM:2012:A69}, \citet{2013MNRAS.435..861R} and \citet{Bagdonaite:2014:10}. 
    These studies uncovered and quantified new systematic errors, such as a varying instrument profile (IP), intra-order wavelength scale distortions, CCD pixel-size irregularities and, more recently, long-range wavelength scale distortions: in their study of possible variations in $\mu$, \citet{2013MNRAS.435..861R} cross-correlated asteroid spectra, observed with VLT--UVES, with laboratory solar spectra and found differential velocity shifts between them of up to 400\,\ms\ over $\sim$600-\AA\ scales. They found these distortions to be an important correction and potential source of systematic error for all previous $\mu$ measurements using VLT--UVES spectra. \citet{Bagdonaite:2014:10} used a similar asteroid technique -- one we describe in detail here -- finding that a similar correction for long-range distortions was required for an accurate measurement of $\mu$ from their quasar spectra. These long-range distortions and the other systematic errors mentioned above have also been factored into more recent \daa measurements \citep[e.g.][]{2013A&A...555A..68M,2013ApJ...778..173E,Evans:2014:128}. However, it remains to be assessed in detail whether these effects may help explain the systematically non-zero \daa measured from the large Keck and VLT samples.

    In this paper, we introduce a new `supercalibration' method of identifying long-range wavelength-scale distortions by using high-resolution echelle spectra of `solar twin' stars and significantly improve the use of asteroid spectra for this purpose.
    We apply this method to HIRES and UVES spectra taken during the epochs contributing to the previous evidence for a varying $\alpha$ and quantify the effect of these errors on previous \daa measurements by modeling the effect on simulated data. 
    The outline of the paper is as follows. 
    In \Sref{sec:observations} we discuss the archival observations and data reductions that have been used in this paper to quantify long-range wavelength miscalibrations.
    In \Sref{sec:methods} we elaborate on the supercalibration method, giving an update to the methods used in \citet{Griest:2009kpa} and \citet{2010ApJ...723...89W}.
    In \Sref{sec:results} we present the results of its application to different spectrographs, and in \Sref{sec:implications} we use simulated spectra to uncover the implications that these wavelength miscalibrations are likely to have had on previous \daa studies.
    We summarize and discuss our findings in \Sref{sec:discussion}.

\section{Asteroid and Stellar Data Reduction} 
\label{sec:observations}
    In this paper we do not consider directly the quasar spectra that led to evidence for a varying $\alpha$ in \citet{Murphy:2003em,Murphy:2004kl} and \citet{2012MNRAS.422.3370K}.
    Instead, we focus on observations using different techniques to quantify the likely types and magnitude of systematic calibration errors in those data.
    Most of the observational data used in this paper for calibration purposes is spectra of either asteroids or stars.
    Our main analysis uses archival, publicly available spectra for both UVES and HIRES across more than 15 years. 
    The spectra were taken across many different nights, sky/telescope positions, weather conditions, temperatures, pressures and spectrograph set-ups.

    \subsection{VLT--UVES} 
    \label{sub:observations_uves}

    The Ultraviolet and Visible Echelle Spectrograph \citep[UVES;][]{Dekker:2000:534} on the Very Large Telescope (VLT) consists of two arms: a blue arm and a red arm.
    The light path when observing an astronomical object is as follows.
    The light reflects off the primary, secondary and tertiary mirrors of the telescope, into a pre-slit unit (containing calibration lamps and optics) where it passes through an image derotator\footnote{``Derotator'' is the name used for the UVES image rotator and we use that convention here.} and then into the Nasmyth-mounted UVES enclosure.
    When a dichroic mirror is in use, the light is split into red and blue light to pass through to the different arms.
    The blue (red) light propagates through the blue (red) slit, a series of optics, the echelle and cross-disperser gratings and is finally imaged onto the CCD. 
    UVES has a total of 3 CCDs, one in the blue arm and two (denoted `upper' and `lower') in the red arm. 
    The ThAr calibration lamp is housed in the UVES enclosure and when a calibration exposure is taken, a mirror is swung into the light path immediately after the telescope shutter, where it reflects the lamp light through the rest of the spectrograph.
    When an iodine cell is being used it is heated and placed in the light path in front of the derotator within the UVES enclosure.

    We use the standard European Southern Observatory (ESO) data reduction software, Common Pipeline Library (CPL) version 4.7.8.
    The reduction scripts are created using UVES\_headsort\footnote{Available at \href{http://astronomy.swin.edu.au/~mmurphy/UVES_headsort/}{\url{http://astronomy.swin.edu.au/~mmurphy/UVES_headsort}}}.
    We also used the carefully-selected ThAr line list found in \citet{Murphy:2007hn} for our reductions and we fit the wavelength solution with a 6th degree polynomial. 
    We discuss the differences and implications of using ``attached'' and ``unattached'' ThAr exposures in \Sref{ssub:results_longrange_uves}.

    \subsection{Keck--HIRES} 
    \label{sub:observations_hires}

    The High Resolution Echelle Spectrograph (HIRES) on the Keck telescope is a cross-dispersed echelle spectrograph with resolving power of $R\approx25000$--85000 \citep{Vogt:1994vh}.
    During an exposure, an astronomical object's light reflects off the primary, secondary, and the tertiary mirrors into the Nasmyth-mounted HIRES enclosure.
    There it passes through the image rotator, through the slit, off the collimator, echelle grating, cross disperser, more mirrors and optics, and is finally imaged onto the CCD.

    The principle aim of this paper is to understand the effect of any wavelength miscalibration on quasar spectra that could lead to errors in previous $\alpha$ measurements. 
    As most of the Keck spectra used in \citet{Murphy:2003em,Murphy:2004kl} were reduced with the reduction software MAKEE, we use MAKEE to reduce the supercalibration spectra throughout this paper. 
    The other main reduction software available is HiresRedux\footnote{Available at \href{http://www.ucolick.org/~xavier/HIRedux/}{\url{http://www.ucolick.org/~xavier/HIRedux/}}} maintained by J.~X.~Prochaska.
    We tested the two software packages by comparing their relative wavelength solutions to the same supercalibration exposures and our tests did not find substantive differences.

\section{Supercalibration Method} 
\label{sec:methods}
    \subsection{General method overview} 
    \label{sub:methods_overview}
    The standard method for wavelength calibrating a quasar exposure on a slit spectrograph is by comparison with a ThAr arc lamp exposure. 
    The calibration source is a ThAr lamp that emits a spectrum with very sharp emission lines at known wavelengths.
    A 2D wavelength solution maps the detector's CCD pixels to wavelengths.
    The ThAr wavelength solution is simply adopted as that of the quasar exposure.
    When accurate wavelength scales are needed, ThAr exposures should clearly be taken before slewing the telescope after the quasar exposure and without changing any of the spectrograph settings.
    However, even when such care is taken, the ThAr light may still follow a different light path through the spectrograph compared to the quasar exposure because the ThAr lamp is typically mounted within the spectrograph.
    And, at the very least, the ThAr light typically illuminates the slit (close to) uniformly, whereas most science objects, e.g., quasars and stars, are point-like, so different point-spread functions are to be expected.
    If those differences are wavelength dependent, a potentially important calibration error will ensue.

    In principle, the accuracy of the ThAr calibration method can be tested by comparing a ThAr-calibrated spectrum with a reference spectrum of the same science object on a different `absolutely calibrated' spectrograph.
    The supercalibration method that we use in this paper follows this basic procedure.
    We take an exposure of a source with the telescope's spectrograph, and compare its final calibrated spectrum with the reference spectrum of the same source from a Fourier Transform Spectrometer (FTS).
    We then solve for a relative velocity shift between these two spectra as a function of wavelength.
    In general, the most useful reference spectra have a large amount of spectral information, i.e., containing many sharp, narrow absorption features.
    We call this the `supercalibration' method: it aims to check the normal calibration against a more reliable reference.

    The basic implementation of the supercalibration method takes advantage of two different reference spectra: a molecular iodine (I$_{2}$) absorption spectrum and the solar spectrum.
    These spectra have a near continuous forest of narrow absorption features that allows us to consider small chunks of each spectrum, typically 500\,\kms wide ($\approx$8\,\AA\ at 5000 \AA).
    We determine the relative velocity shift between the reference chunk and the object spectrum using the following five transformations to each reference chunk: 1) a constant wavelength shift, 2) a flux scaling factor, 3) a constant flux offset, 4) a linear continuum slope correction, and 5) the velocity width of a Gaussian instrument profile (IP).
    Because we are solving for transformations of the entire spectrum within each chunk we are using all of the spectral information available. 
    We solve for the best-fitting model parameters by minimizing $\chi^2$ between the transformed reference chunk and the object spectrum using \textsc{iminuit}, the \textsc{python} implementation of \textsc{seal minuit}\footnote{Available at \href{http://iminuit.github.io/iminuit/}{\url{http://iminuit.github.io/iminuit/}}}.
    
    In all that follows, the reference will be a high \SNR FTS spectrum of the relevant reference source, either the solar spectrum or the iodine cell absorption spectrum.
    The FTS is an instrument that works under a fundamentally different process than the echelle spectrographs on telescopes, and we expect the relative FTS frequency scale (and thus wavelength scale) to be much more accurate.
    Therefore, the reference FTS spectrum's wavelength scale is assumed to be correct; this assumption can be relaxed in certain ways that are detailed later (see \Sref{ssub:methods_solar_test}).
    The best-fit wavelength shift for each chunk is given by the following relation: $\lambda_{\rm{shift}} = \lambda_{\rm{FTS}} - \lambda_{\rm{ThAr}}$.
    We plot our results in this paper with this wavelength shift turned into a velocity shift via
    \begin{equation}
        v_{\rm{shift}} = c \times \frac{\lambda_{\rm{shift}}}{\lambda}.
    \label{eq:vshift}
    \end{equation}
    The final, reported shift should be added to the standard ThAr calibration solution to obtain the `correct' wavelength, i.e., that matching the FTS reference. 
    For example, a \vshift for a given chunk of +50 \ms means that the science spectrum needs 50 \ms added to its wavelength scale for that chunk to align with the reference spectrum.

    We note two practical details here about implementing the supercalibration method before describing our two implementations in the following two subsections.
    First, for all the analysis that follows, we mask atmospheric lines by removing the portion of the spectrum around and including the wavelength of known telluric emission and absorption lines\footnote{Available at \href{http://www.astro.caltech.edu/~tb/makee/OutputFiles/skylines.txt}{\url{http://www.astro.caltech.edu/~tb/makee/OutputFiles/skylines.txt}}}.
    The presence of strong atmospheric lines can cause a spurious value for the \vshift derived from the supercalibration method.
    Second, our method is applied to each echelle order separately and not on order-merged or redispersed exposures, so no artifacts or spurious velocity shifts can enter from redispersing or combining overlapping edges of adjacent orders.
 
    \subsection{Supercalibration method: iodine cell implementation} 
    \label{sub:methods_iodine}
    The UVES and HIRES instruments each contain a glass enclosure containing molecular iodine gas (I$_{2}$) called an iodine cell.
    This is heated to either 70$^{\circ}$C (UVES) or 50$^{\circ}$C (HIRES) and, at the observers' request, placed directly in the light path of the spectrograph.
    As light passes through the iodine cell a forest of narrow absorption features is imprinted from roughly 5000--6200\,\AA.
    The iodine cell has been used extensively for exoplanet detection \citep{1995PASP..107..966V,1996PASP..108..500B,Bedding:2001fa,Butler:2003fq,Marcy:2005jk}. 

    We make use of the iodine cell as an independent check of the ThAr wavelength solution.
    The first implementation of this method used the iodine cell spectra imprinted onto a quasar spectrum \citep[see][for details]{Griest:2009kpa,2010ApJ...723...89W}, and we use it again for the current work with a few minor improvements (allowing a constant flux offset, for example).
    The main difference is that here we use a fast-rotating bright star as the source because these are much brighter than quasars.
    The rotation effectively blurs any sharp spectral features from the star itself.
    After calibrating this exposure with a ThAr exposure in the standard way, we use the supercalibration method to compare it with a laboratory FTS spectrum of the iodine cell as the reference. 
    
\begin{figure}
        \includegraphics[width=1.0\columnwidth]{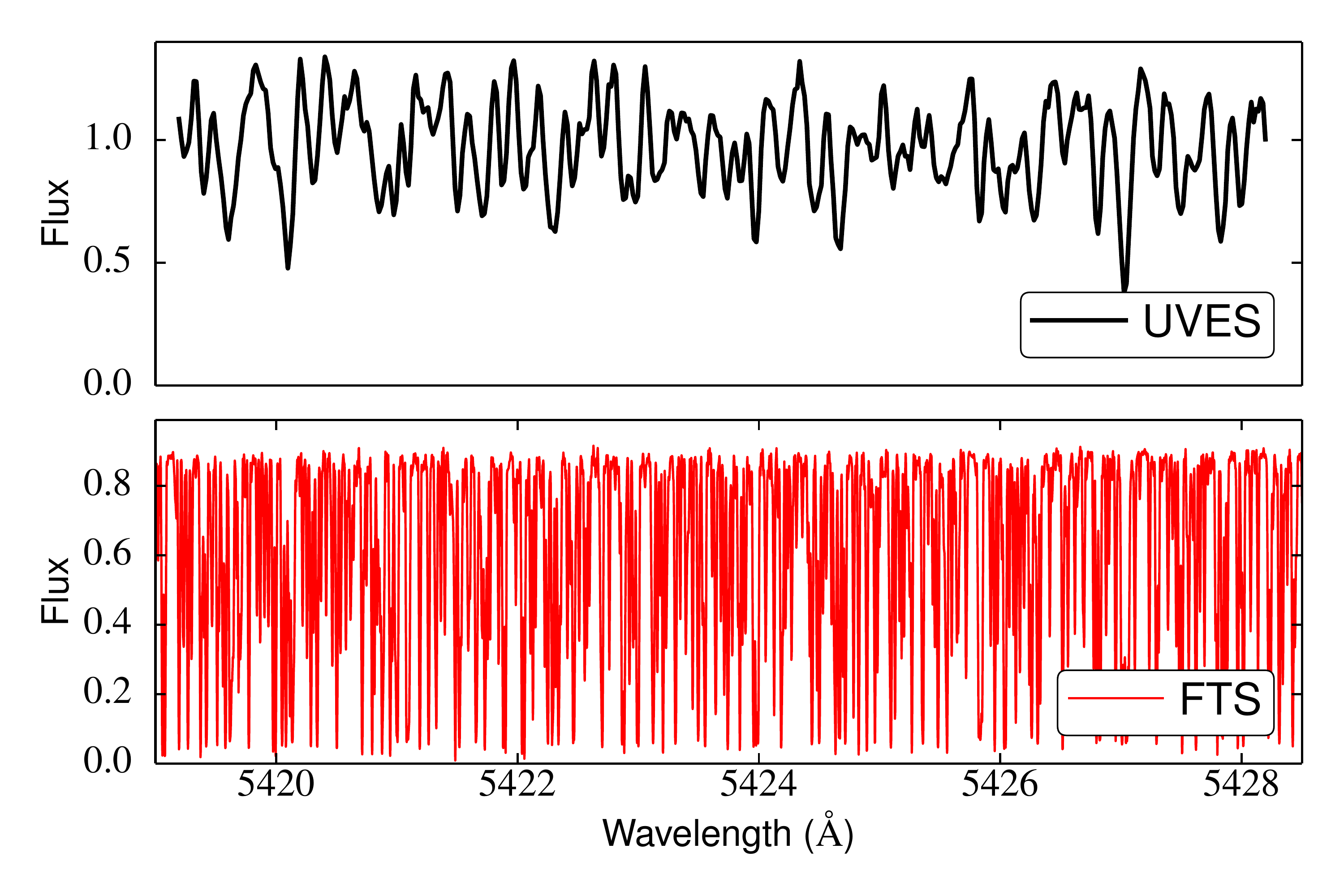}\vspace{-1.0em}
\includegraphics[width=1.0\columnwidth]{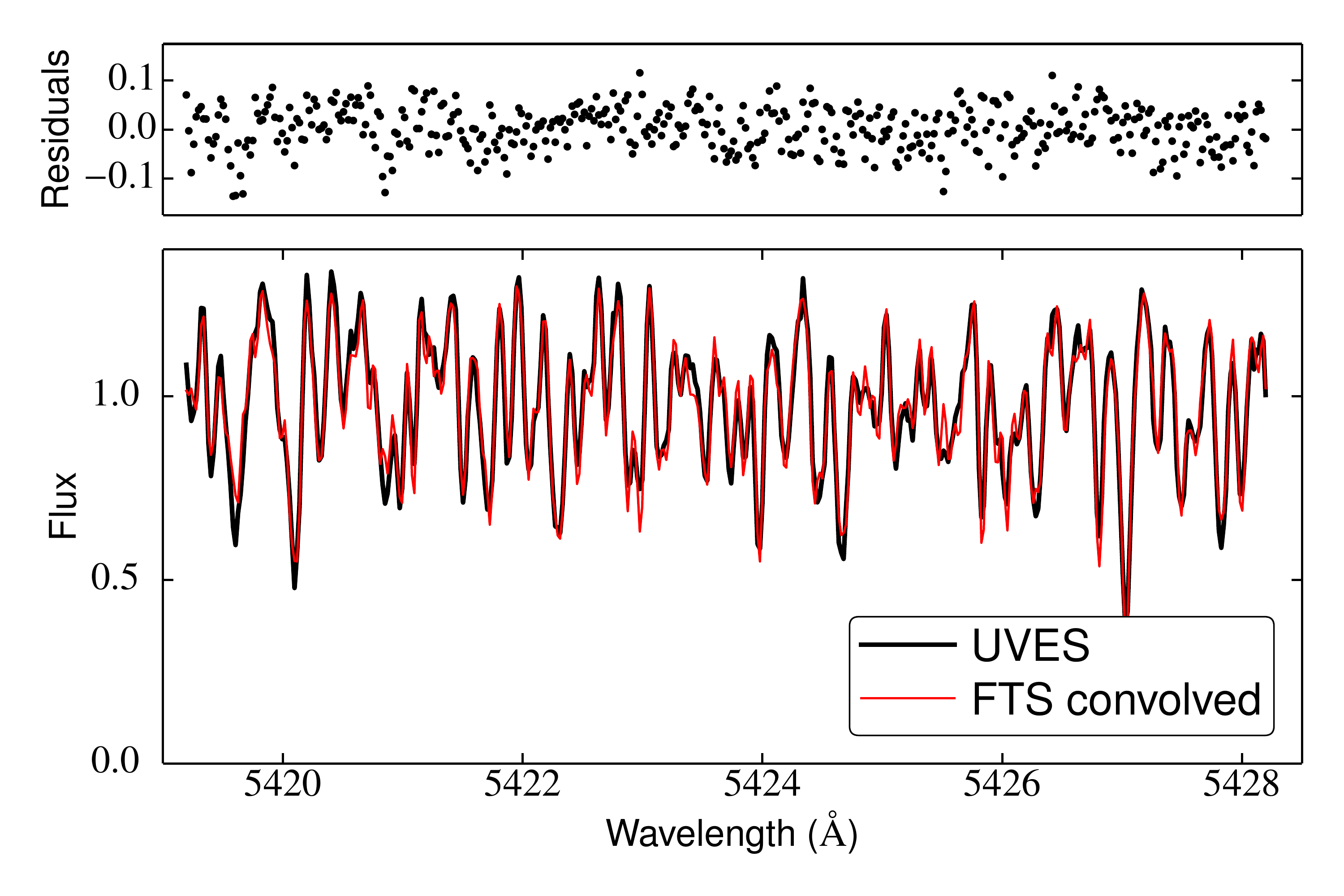}\vspace{-1.5em}
        \caption{An example wavelength chunk of the iodine cell supercalibration method.
        The top panel shows the iodine cell spectrum imprinted on the spectrum of HR9087, a fast-rotating bright star, observed with the UVES spectrograph on December 9, 2009.
        The second panel shows the same region of the iodine cell spectrum as measured by a Fourier Transform Spectrometer (FTS).
        The third panel shows the normalized residuals of the best-fit `supercalibration' for the chunk shown in the fourth panel.}
        \label{fig:best-single-bin-iodine}
    \end{figure}

    The FTS spectrum\footnote{UVES iodine cell FTS scan available from ESO at \href{http://www.eso.org/sci/facilities/paranal/instruments/uves/tools.html}{\url{http://www.eso.org/sci/facilities/paranal/instruments/uves/tools.html}}} of the iodine cell has much higher resolving power ($R\approx800,000$) than the telescope echelle spectrographs, UVES or HIRES ($R\approx50,000$--80,000), used in the relevant quasar studies.
    \Fref{fig:best-single-bin-iodine} shows the iodine cell reference spectrum with the corresponding telescope spectrum.
    The much higher resolving power of the FTS is clearly visible. 
    During the fitting process, each ($\approx$8\,\AA) chunk of the FTS spectrum is transformed via the free parameters listed in \Sref{sub:methods_overview} until $\chi^2$ is minimized.
    \Fref{fig:best-single-bin-iodine} shows an example best-fit solution.
    The best-fit transformed FTS spectrum produces the bottom panel in \Fref{fig:best-single-bin-iodine} and clearly shows the good match between the model and the data.
    We use the iodine supercalibration method in \Sref{sub:results_sliteffects}, where we characterize the effect of object placement within the spectrograph slit. 
    
    \subsection{Supercalibration method: solar implementation} 
    \label{sub:methods_solar}
    The main results of this paper derive from using the supercalibration method with the solar spectrum.
    We extend the same supercalibration method used by the iodine cell, but instead of iodine, we use the solar atmosphere as the reference source.
    The main advantage of using the solar spectrum is that sharp, narrow absorption lines are present at all optical wavelengths, not just at 5000--6200\,\AA\ as in the iodine method.
    Another advantage is that no change in the telescope's optical path is required.
    In contrast, after placing the iodine cell in the light path, adjustments usually have to be made to the focus and slit alignment.
    The solar FTS reference spectrum we use comes from the Chance/Kurucz FTS vacuum solar spectrum \citet{2010JQSRT.111.1289C}, which they refer to as KPNO2010\footnote{Available at \href{http://kurucz.harvard.edu/sun/irradiance2005/irradthu.dat}{\url{http://kurucz.harvard.edu/sun/irradiance2005/irradthu.dat}}}.
    To access the solar spectrum at the telescope we observe two types of astronomical objects: asteroids and `solar twins'.
    The first uses the Sun's spectrum itself reflected off solar system objects, in this case asteroids (moons are another possibility).
    The second uses stars that have almost identical spectra to our Sun and, for the intents and purposes here, can be treated as real solar spectra.
    We present the latter method here for the first time, demonstrating that it is effective and has several advantages over the other methods. 

    \subsubsection{Asteroid supercalibration} 
    \label{ssub:methods_solar_asteroid}
    
    The asteroid-reflected solar spectrum has been used before for spectrograph calibrations.
    \citet{Molaro:2008hc} first fitted for a relative radial velocity difference between the centroids of a small number of selected, individual absorption lines and others in the same spectrum and also with the corresponding lines in a laboratory FTS solar spectrum. \citet{2013MNRAS.435..861R} instead used a cross-correlation technique to compare spectral regions approximately one echelle order wide ($\approx$80\,\AA), containing many absorption lines, between asteroid and laboratory FTS spectra.
    Our technique uses small spectral chunks, 500\,\kms\ ($\approx$8\,\AA) wide, which allows us to map wavelength calibration distortions not only over long wavelength ranges, but also in detail across each echelle order.

    \subsubsection{Solar twin supercalibration} 
    \label{ssub:methods_solar_solartwin}
\begin{figure}
        \includegraphics[width=1.0\columnwidth]{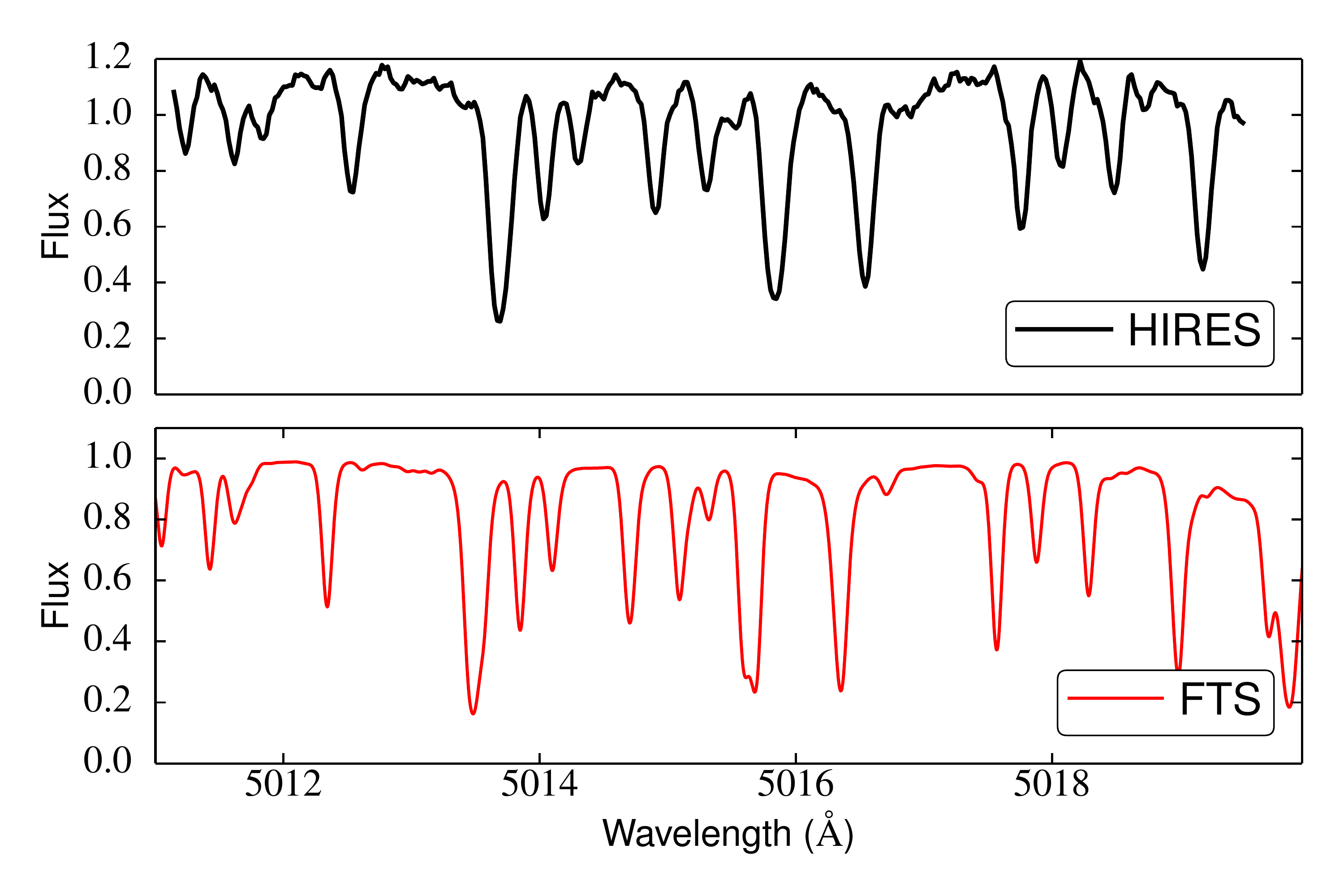}\vspace{-1.0em}
\includegraphics[width=1.0\columnwidth]{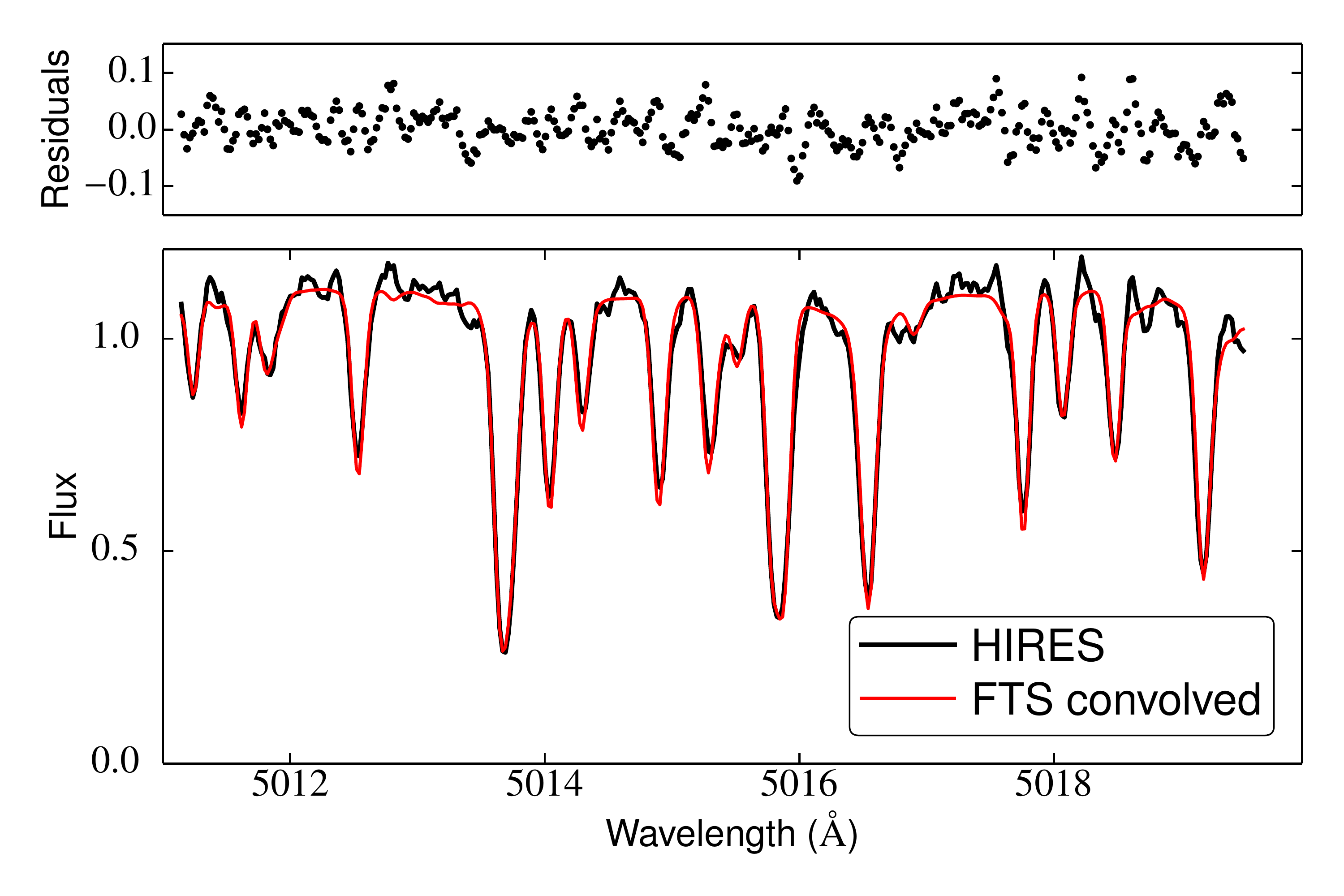}\vspace{-1.5em}
        \caption{An example wavelength chunk of the solar twin supercalibration method.
        The top panels shows the spectrum of HD146233, a solar twin, observed with HIRES on July 13, 2007.
        The second panel shows the same wavelength region of the solar spectrum as measured by a Fourier Transform Spectrometer (FTS).
        The third panel shows normalized residuals of the best-fit `supercalibration' for the chunk shown in the fourth panel.}
        \label{fig:best-single-bin}
    \end{figure}

    We present a new technique for comparing spectra of `solar twin' stars and solar FTS spectra.
    This technique is the same as for the asteroids in all respects (same reference FTS, same velocity chunk size, etc.), except that it uses a solar twin as the object spectrum, rather than solar light reflected from solar system objects.
    In this work we restricted observations to stars defined as `twins' or `analogues' of the Sun in \citet{Takeda:2007:663}, \citet{Melendez:2009:L66}, \citet{Onehag:2011:A85} and \citet{Datson:2014:1028}. However, we have recently observed other `solar-like' stars -- those with similar spectral types to our Sun, e.g.~G1V--G4V -- and compared their supercalibration results with those from known solar twins; the results are very similar and it may well be possible to obtain reliable supercalibration information from a much larger number of stars than just those deemed as solar `twins' or `analogues'.
    The close match between the solar FTS and typical solar twin spectra from the telescope can be easily seen in~\Fref{fig:best-single-bin}.
    The number density of spectral features in the solar spectrum is such that the solar twin (and asteroid) supercalibration technique provides a velocity shift measurement with statistical precision of $\approx$12\,\ms\ for each 500-\kms\ chunk when the spectral \SNR is $\approx$120\,per 1.3-\kms\ pixel. With $\approx$10 chunks per echelle order, this is adequate for diagnosing relative velocity distortions of $\ga$30\,\ms\ over intra-order and/or long-range wavelength scales. This statistical error derives only from the photon noise in the spectra, so it will diminish as \SNR$^{-1}$. However, it will only improve with increasing resolving power if the solar absorption lines are individually unresolved and not substantially blended together. \Fref{fig:best-single-bin} illustrates that, at least at resolving powers typical of most quasar observations ($R\sim50000$), the latter is not generally true. Therefore, the statistical precision of this method is unlikely to improve substantially with increasing resolving power.

    Several concerns can be raised about the use of solar twins.
    First of all, the absorption features of Sun-like stars change, so individual absorption lines in the FTS reference and the solar twin spectrum may have slightly different relative optical depths.
    However, the way we derive the supercalibration information is not from any particular line, but from chunks of a spectrum which contain many lines.
    And, even if relative optical depth changes to many lines conspired to produce a spurious velocity shift in one chunk, the effect will be different in other chunks. 

    Second, it is important to realize that, even if the solar twin is not exactly the same as our Sun and the relative strengths of some lines differ, the underlying transition frequencies in the two spectra are the same. Despite this, stellar activity, including the motions of convective cells and starspots, will cause shifts between the centroids of the absorption features in the star relative to the FTS solar spectrum. These effects will have both random and systematic components which are much less than the typical line width, i.e.~$\la$100\,\ms. Line-to-line random shifts are diminished in our technique because each chunk includes many lines, while the systematic component should not be important for determining wavelength calibration distortions (i.e.~systematic velocity shift variations with wavelength). Due to similar activity variations in our own Sun, our asteroid technique will also suffer from some of these effects. However, a line-by-line analysis, like that first conducted by \citet{Molaro:2008hc}, coupled with monitoring of solar activity, can in principle remove these effects.

    Finally, if wavelength calibration distortions are driven by drifts in the spectrographs, it may be a problem that there is a substantial difference in the exposure times between solar twins and quasars. 
    However, the benefit of a short exposure time is that the calibration check is relatively rapid and can be taken during the night.
    A possible advantage of this technique over using reflected solar light from solar system objects (e.g. asteroids like in \Sref{ssub:methods_solar_asteroid}, or the moon) or sky emission spectra \citep[e.g.][]{1995PASP..107..966V}, is that stars are unresolved point sources\footnote{Some bright asteroids project angular sizes of $\la$0\farcs2 which, for the purposes of most optical observations, where the seeing is typically $\ga$0\farcs5, is unresolved. However, the brightest asteroids (e.g.~Ceres and Vesta) can be marginally resolved in such conditions.}.
    This means that they offer the closest comparison to the quasar observations [aside from directly comparing two spectra of the same object, which can provide information about relative calibration distortions, e.g.~\citet{2013ApJ...778..173E}].
    Solar twins are also at fixed positions in the sky -- they do not move at substantially non-sidereal rates like asteroids and other solar system objects. This increases the practicality of solar twin supercalibration.

    \subsubsection{Test of solar FTS spectrum} 
    \label{ssub:methods_solar_test}

    Another possible concern about using the solar FTS reference is the assumption that its wavelength scale is correct and does not suffer from any systematic distortions itself. 
    The FTS spectrum was constructed by stitching together several overlapping FTS scans of the solar spectrum \citep{2010JQSRT.111.1289C}, and this stitching could introduce long-range wavelength distortions. 
    A few of these concerns can be addressed by some relatively simple checks. 

    The High Accuracy Radial velocity Planet Searcher (HARPS) spectrograph is a very stable, well-calibrated, temperature controlled and, most importantly for the present considerations, fibre-fed vacuum spectrograph on the ESO La Silla 3.6m telescope \citep{Mayor:2003wv}. 
    It differs in a number of ways from a slit spectrograph in air (UVES and HIRES): the optical fibre feeds the light directly to the instrument, the instrument is held to a constant temperature, a second fibre can constantly input the ThAr spectrum to monitor any potential change in the wavelength scale, the instrument is kept in a vacuum, and the fibre scrambles the image before it is fed into the instrument. 
    Nevertheless, studies of HARPS using highly accurate laser frequency-comb calibration by \citet{Wilken:2010fc} and \citet{2013A&A...560A..61M} indicate that ThAr calibration results in $\approx$70\,\ms distortions on echelle order scales ($\approx$100\,\AA). The intra-order distortions were proposed to arise from pixel size changes near the edges of the 512-pixel manufacturing stamp pattern and \citet{2013A&A...560A..61M} ruled out errors in the laboratory wavelengths of the ThAr lines as another possible origin. Neither study revealed evidence for long-range distortions, though they covered relatively short/moderate wavelengths ranges ($\sim$50\,\AA\ at 5150\,\AA\ and 4800--5800\,\AA, respectively). Nor are long-range distortions evident in the results of the line-by-line comparison by \citet{Molaro:2011:A74} between HARPS asteroid spectra and solar altases \citep{AllendePrieto:1998:41,AllendePrieto:1998:431} at 4000--4100 and 5400--6900\,\AA.

\begin{figure}
        \includegraphics[width=1.0\columnwidth]{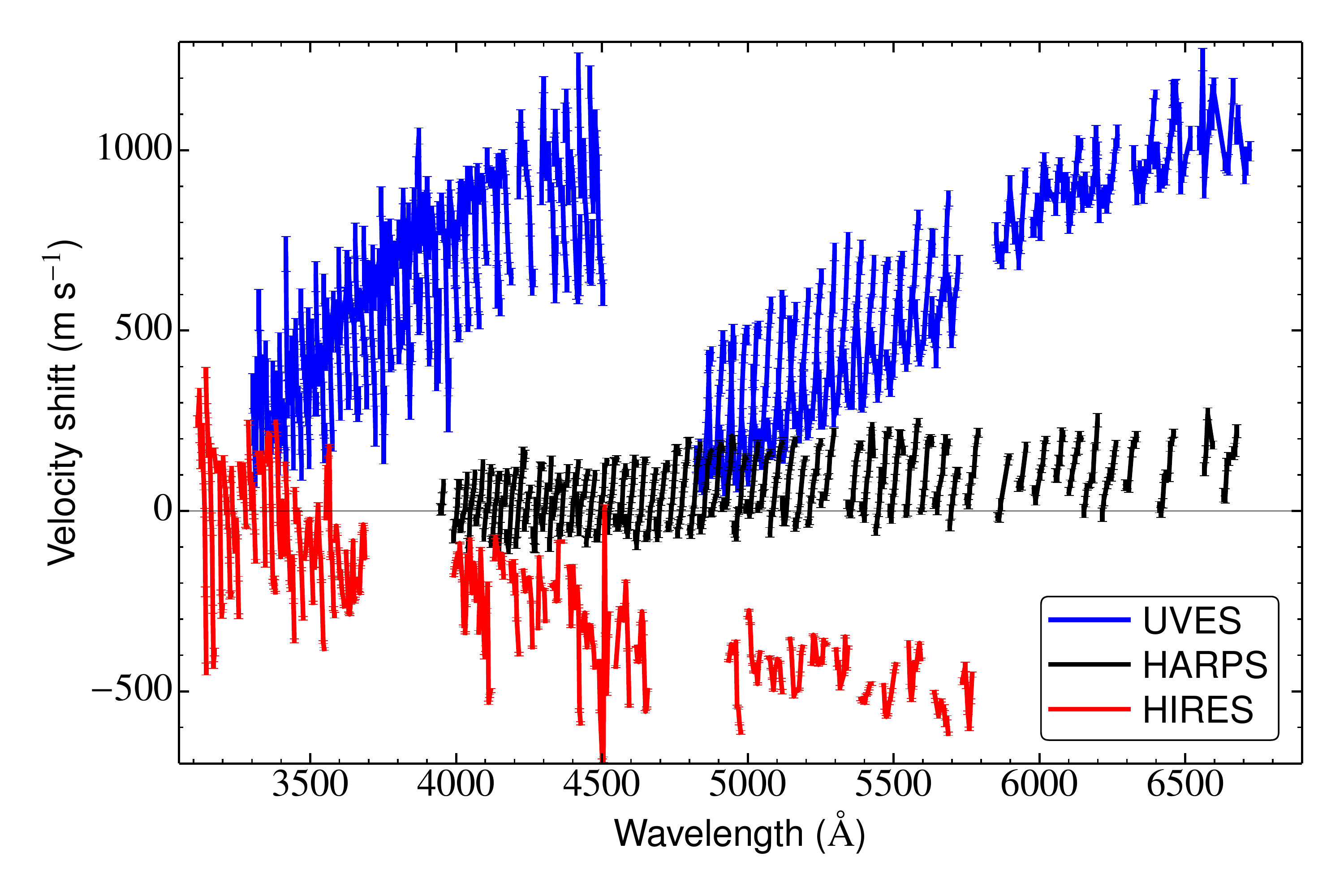}\vspace{-1.5em}
        \caption{The solar twin supercalibration of three different spectrographs compared on a single scale. 
        The vertical offsets of each spectrograph are shifted by an arbitrary constant velocity so that the overall structure can be easily compared.
        HIRES and UVES are the slit spectrographs analyzed in detail in this paper, while HARPS is a fibre-fed and extremely stable spectrograph shown for comparison.
        The long-range distortions of the slit spectrographs in comparison to the fibre-fed spectrograph are clearly visible as non-zero slopes.
        The velocity shift here and throughout this paper is defined according to \Eref{eq:vshift}; the velocity shift should be added to the ThAr wavelength to match the FTS reference.
        \label{fig:uves-hires-harps}}
    \end{figure}

    The results of a solar twin supercalibration using HARPS is shown in \Fref{fig:uves-hires-harps}. We see clear evidence intra-order distortions in these HARPS results, but with very little long-range wavelength calibration distortion. The intra-order distortions are similar in both shape and magnitude to those seen in the laser frequency-comb calbibrated HARPS spectra of \citet{Wilken:2010fc} and \citet{2013A&A...560A..61M}. This confirms that the solar twin supercalibration technique returns reasonably reliable information about relative distortions. However, our results are not identical to the frequency-comb results: our intra-order distortions typically span approximately $\pm100$\,\ms\ across each order, with either a flattening or slight reversal of the intra-order slope towards the order edges, whereas the results of \citet{Wilken:2010fc} and \citet{2013A&A...560A..61M} generally show a $\pm70$\,\ms\ span with the slope reversing completely towards the order edges. An important possibility is that the intra-order distortions are somewhat variable within HARPS, so further testing of different supercalibration techniques with HARPS is desirable. There may be some evidence in \Fref{fig:uves-hires-harps} for a small, $\sim$45\,\msperthousand long-range distortion in the HARPS results and/or a $\sim$50\,\ms\ shift at $\sim$4750\,\AA. Given the lack of evidence for long-range slopes of this magnitude in the previous HARPS asteroid tests \citep{Molaro:2011:A74,2013A&A...560A..61M}, it may be that the slope and/or shift we observe is due to systematic effects in the \citet{2010JQSRT.111.1289C} solar FTS spectrum. We do not attempt to correct this possible effect in the results of this paper because it is significantly smaller than the typical distortions we find in the UVES and HIRES instruments. Nevertheless, it highlights the importance of obtaining a more accurate solar reference spectrum for use in supercalibrating astronomical spectrographs.

    \Fref{fig:uves-hires-harps} also shows an example solar twin supercalibration from both UVES and HIRES. The relative differences between the results from these slit spectrographs and the fibre-fed HARPS instrument is striking and compelling.
    There appear to be substantial long-range wavelength calibration distortions in both UVES and HIRES.
    Interestingly, it appears that each arm of UVES suffers from a long-range distortion with a similar slope, while the single arm of HIRES is characterized by a single distortion.
    The supercalibration for each of these three instruments uses the same FTS reference, so any intrinsic distortions from the supercalibration method itself must be much smaller than the effect we are detecting in UVES and HIRES.
    Finally, even if there is some conspiratorial way that the FTS and HARPS spectrographs have wavelength distortions that effectively cancel each other, the relative differences between HIRES and UVES still remain (though their absolute levels of distortion would be unknown).

    In summary, the solar twin supercalibration is a new approach that reliably exposes and quantifies distortions in the ThAr wavelength solution. The next section details the different long-range wavelength scale distortions in both the UVES and HIRES instruments. 

\section{Velocity Distortions in VLT--UVES and Keck--HIRES} 
\label{sec:results}
    
    One key advantage of the solar twin supercalibration method is that exposures of various solar twin stars have been taken with Keck--HIRES and VLT--UVES over many years.
    Applying the solar twin supercalibration method to both the Keck--HIRES and VLT--UVES archival exposures, taken across a wide range of dates and conditions, therefore allows us to quantify the historical record of long-range velocity distortions of the instruments.
    Below, we find significant long-range velocity distortions in both Keck--HIRES and VLT--UVES over most of their lifetimes, albeit somewhat sparsely sampled.
    We examine the distortions themselves in this section, while we analyze the impact that these distortions may have had on previous fundamental constant measurements in \Sref{sec:implications}.

    \subsection{Long-range velocity distortions}
    \label{sub:results_longrange}

    \subsubsection{VLT--UVES} 
    \label{ssub:results_longrange_uves}

    As described in \Sref{sub:observations_uves}, the UVES spectrograph has the option of using a dichroic mirror to split incoming light into two arms (red and blue).
    The red arm consists of two CCDs, while the blue arm has a single CCD.
    There are a number of standard wavelength settings for UVES that are referred to by their central wavelength in nm for the two arms.
    For example, the 390/580 setting centres the blue arm at 3900\,\AA\ and the red arm at 5800\,\AA.
    The central wavelength of the blue arm falls in the centre of the blue CCD, while in the red arm the central wavelength falls between the two CCDs.

    UVES was designed so that the gratings could be reliably set, changed, and reset to the same position such that dispersed wavelengths were directed to the same CCD location to within $\approx$1/10$^{th}$ of a pixel.
    The goal of this design was to maximize time during the night spent taking science exposures, with the ThAr calibration exposures to be taken at the end of the night. 
    Because of this design, the gratings are automatically reset after each observation block (OB) during the standard observation protocol.
    ThAr exposures taken in this way are referred to as ``unattached''.
    However, OBs may include an ``attached'' calibration exposure in which the science exposure is followed immediately by a ThAr exposure without any intervening grating reset.
    Such an ``attached'' ThAr is clearly preferred when accurate wavelength calibrations are required.
    Note that the spectra used by \citet{2012MNRAS.422.3370K} in the larger VLT quasar sample of \daa measurements were taken predominantly in the ``unattached'' mode.

\begin{figure}
        \includegraphics[width=1.0\columnwidth]{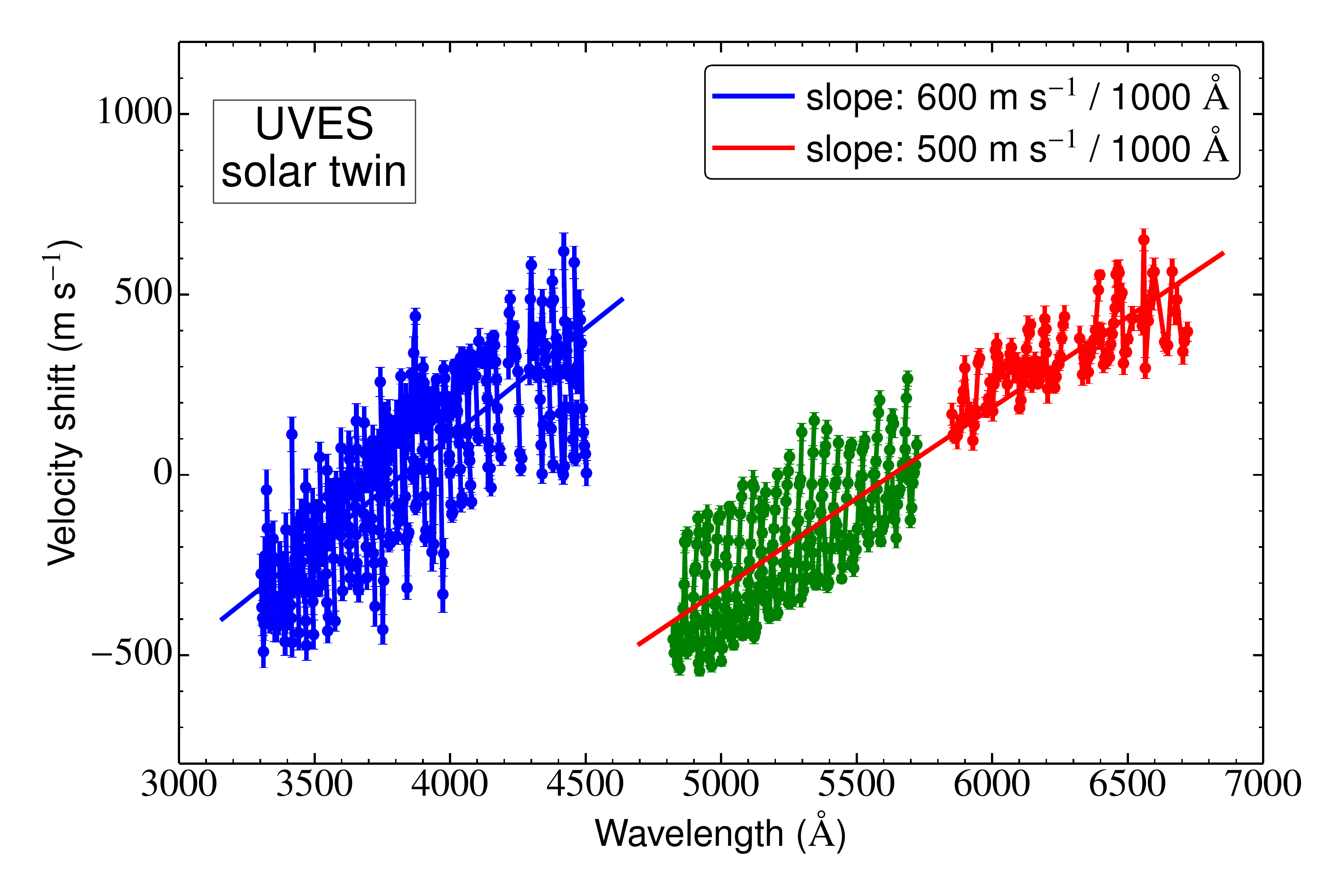}\vspace{-1.0em}
        \includegraphics[width=1.0\columnwidth]{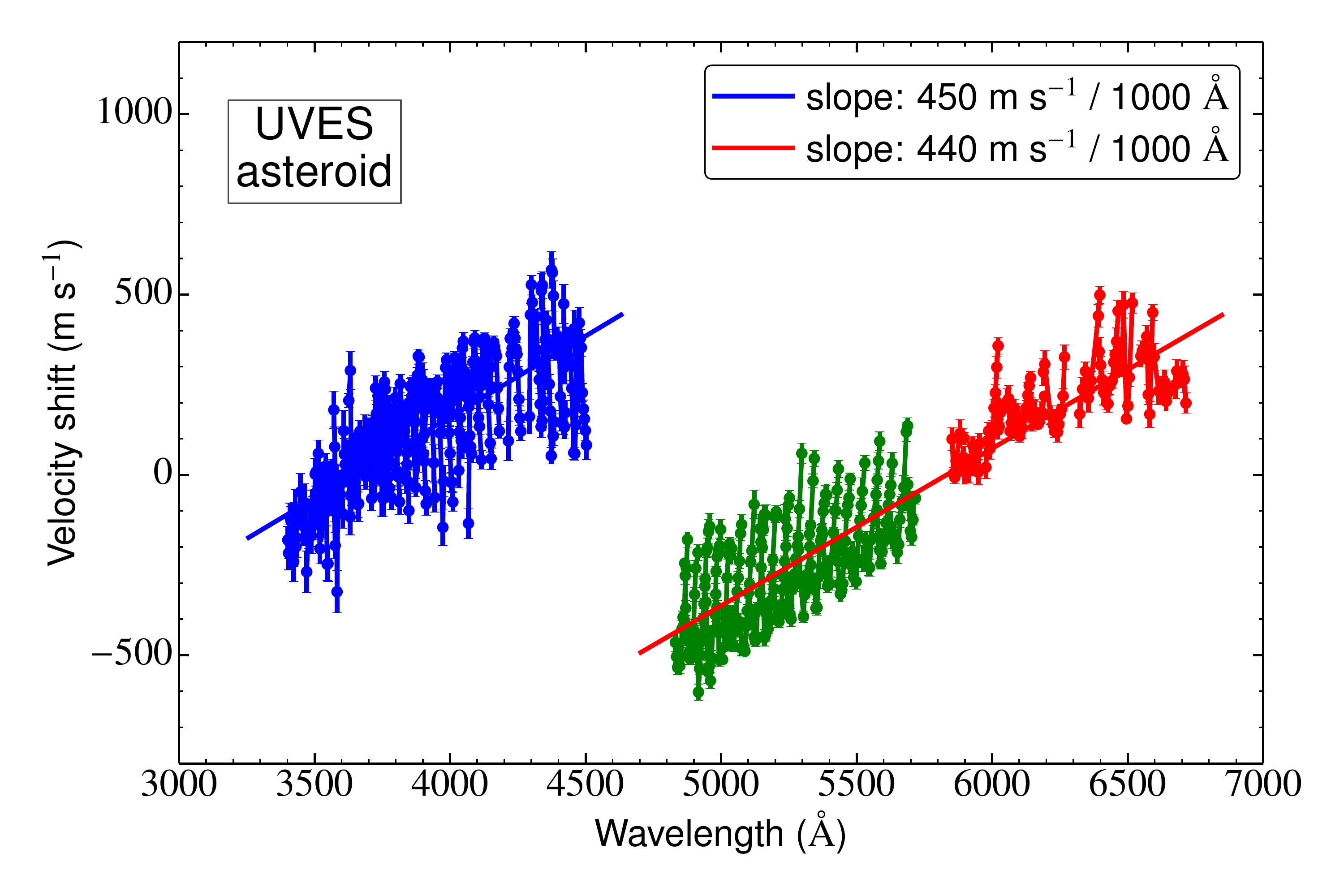}\vspace{-1.5em}
        \caption{The supercalibration of solar twin HD76440 (upper) and asteroid Eunomia (lower) observed with UVES on May 15, 2013.
        The blue CCD is plotted in blue (left), the two red CCDs are plotted in green (middle) and red (right).
        In each exposure, the same arbitrary velocity shift has been applied to the results of all 3 CCDs so that the results roughly centre on zero velocity shift.
        \label{fig:uves-distortion}}
    \end{figure}

    On 15th of May, 2013, the solar twin HD76440 and the asteroid Eunomia were observed several times, each with an attached ThAr exposure, over a 30 minute period.
    The telescope was slewed between these exposures.  
    We plot the resulting supercalibration method results for both objects in \Fref{fig:uves-distortion}. 
    Intra-order (across a single echelle order) velocity distortions are prevalent in all orders and an approximately linear long-range velocity distortion appears across the wavelength range of each arm separately.
    The long-range velocity slope is found by fitting an unweighted line to the average \vshift\ per order for each arm. 
    There appears to be a similar slope in each arm such that the total velocity shift over the wavelength range of each arm is approximately the same.
    There also appears to be little or no velocity shift between the central wavelengths of the two arms.
    We find this to be a common distortion pattern for UVES and, from the archival data we have, find that it appears to be independent of the wavelength setting used in each arm.
    Because the centre of the arms are aligned in velocity space, at least in these particular exposures, the common slope effectively translates into an overall ``lightning-bolt'' shaped distortion across the wavelength coverage of the exposure.

    We have sparse archival data that irregularly samples the wavelength miscalibrations over the time range of the quasar observations we consider.
    Nonetheless, we find a few overall characteristics that are shared generally by most of the supercalibration exposures we analyzed.
    The slope of the distortions appear to be of similar slope and sign between the two arms, i.e. if the blue arm has a high positive slope, the red arm will also have a high positive slope.
    We also find that positive slopes are more common in the archival supercalibration spectra we were able to identify.
    To plot this slope we adopt the following simple procedure.
    A weighted average \vshift for each order was first calculated, then an unweighted straight line was fit to each average order value within each arm of the spectrograph. 

\begin{figure}
        \includegraphics[width=1.0\columnwidth]{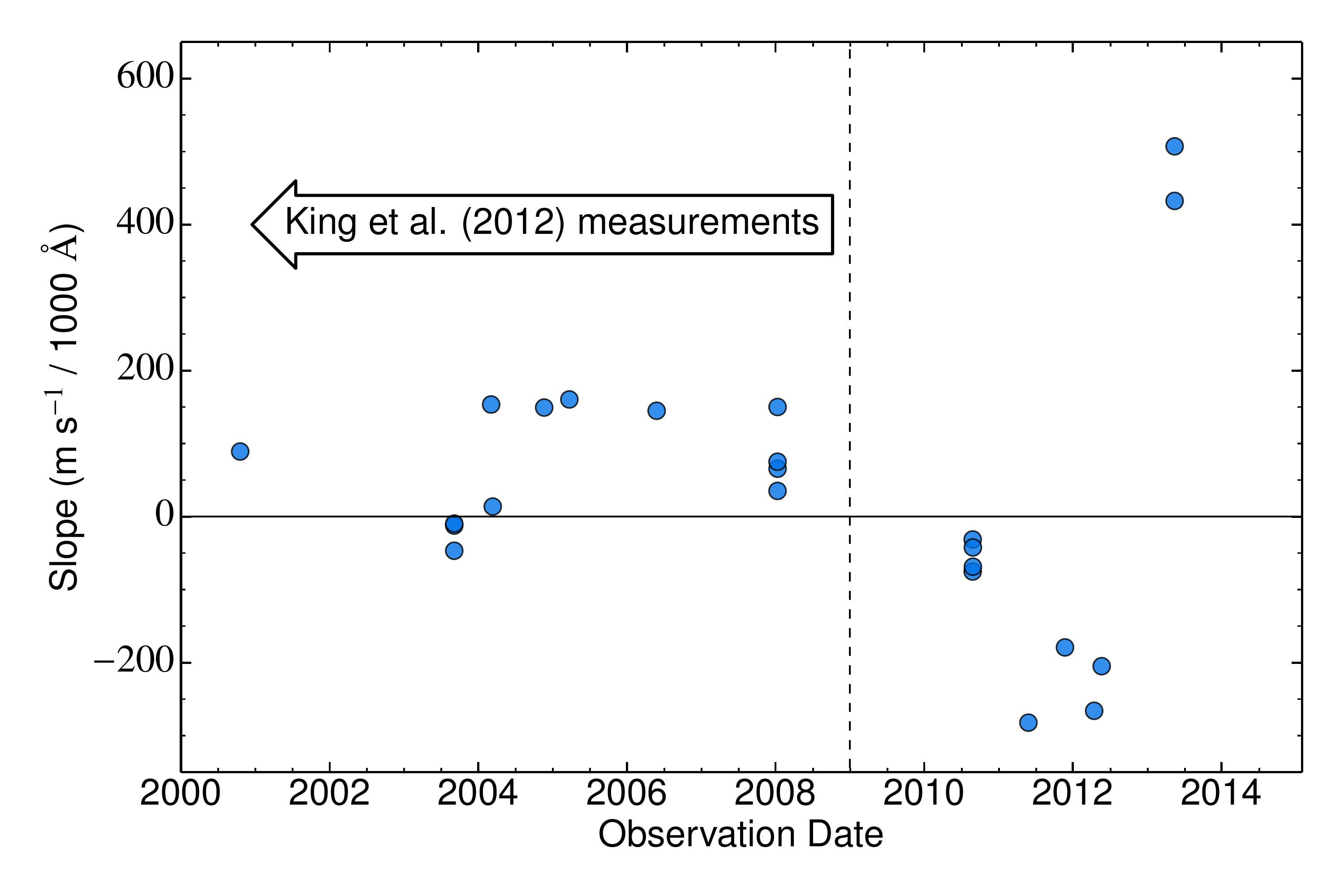}\vspace{-1.5em}
        \caption{The best-fit supercalibration slope of each arm in UVES for all archival solar twin and asteroid exposures we could identify with spectrograph setups similar to those used for the quasar spectra used by \citet{2012MNRAS.422.3370K}.
        The period before the vertical line denotes that in which the \citeauthor{2012MNRAS.422.3370K} quasar spectra were observed.}
        \label{fig:uves-slope-history}
    \end{figure}

    As a visual summary of this best-fit linear slope to the long-range velocity distortions, we plot the slopes found as a function of observation date in \Fref{fig:uves-slope-history}.
    This figure shows the results of the supercalibration analysis on the solar spectra found in the UVES archive with slit widths and settings that were also used during quasar observations.
    As is clearly seen, there is a fairly wide range of long-range velocity slopes across the sparsely sampled history of UVES.
    The range of slopes is typically between $\pm$200\,\msperthousand with increased divergence after 2010.
    Data for the \citet{2012MNRAS.422.3370K} were all taken in the years before 2009.
    The different slopes over time do not suggest any simple characterization as a function of time.

    A number of hypotheses were tested in an attempt to correlate the long-range slopes and parameters in the header values of the object exposures as well as the corresponding ThAr exposures (e.g.~seeing, telescope altitude/elevation, slit-width, etc.), with no clear relationship found. There are some hints that the distortions are caused by a combination of effects, with at least one potentially deterministic cause investigated in \Sref{sub:ssub:results_uvesstability}.
    \Fref{fig:uves-slope-history} may also hint that the spectrograph's wavelength distortions might be quasi-stable for a period of time before an event (e.g.~a mechanical realignment or earthquake) creates a larger change. An example of this may be the period $\sim$2004--2008 over which the distortion slope appears to be $\sim$200\,\msperthousand. However, a similar slope does not seem apparent in the asteroid observations from 2006 December in \citet{Molaro:2008hc}. Also, \citet{2013MNRAS.435..861R} found substantially different slopes in asteroid exposures from 2006, 2010, 2011 and 2012 ($\sim$315, 130, 210 and 615\,\msperthousand, respectively) for the blue arm of UVES compared to our supercalibration results in the red arm in the same years. Therefore, the lack of slope variations seen in \Fref{fig:uves-slope-history} for $\sim$2004--2008 may well be due to the very sparse sampling available.

    Finally, in principle the supercalibration approach allows us to check for and measure any overall velocity shifts between the two arms of UVES. Obviously, such shifts might arise from misalignment of the two corresponding entrance slits, as projected on the sky, but they might also arise from the same (as yet unknown) physical effect that causes the long-range distortions. \citet{Molaro:2008hc} found that the UVES `arm shift' was $\la$30\,\ms\ in their asteroid observations. For the 2013 solar-twin and asteroid results shown in \Fref{fig:uves-distortion}, there is also no clear velocity offset between the blue and red arms of UVES. However, the archival solar-twin supercalibrations provide few opportunities to measure the offset: in all but 4 cases, the ThAr wavelength calibration exposures for the blue and red arms were not taken simultaneously in a single dichroic setting like the stellar exposure. Wavelength setting changes between the blue- and red-arm ThAr exposures should be expected to produce velocity shifts of up to $\sim$300\,\ms\ in those cases. From the 4 exposures in which appropriate inter-arm calibration is available (in mid-2010 and mid-2013), we find velocity shifts $<$100\,\ms\ by comparing the fitted slopes at the nominal wavelength centres of the two arms.

    \subsubsection{Keck--HIRES} 
    \label{ssub:ssub:results_longrange_hires}

\begin{figure}
        \includegraphics[width=1.0\columnwidth]{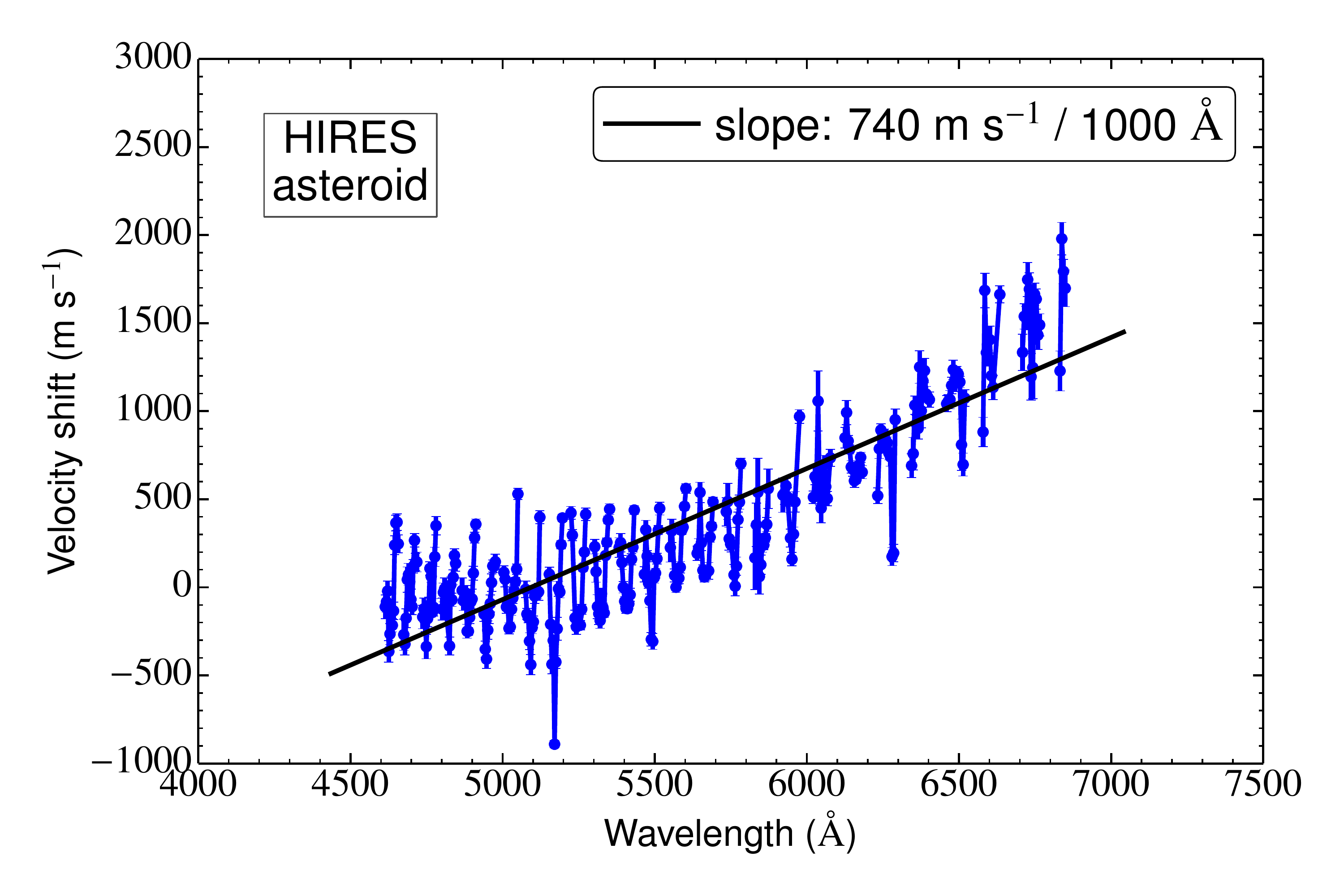}\vspace{-1.5em}
        \caption{The supercalibration result for an exposure of Ceres (asteroid) taken with HIRES on July 5, 1995.
        The exposure was taken in the ``pre-image rotator'' era of HIRES. 
        The resulting supercalibration velocity shift is plotted after an arbitrary constant velocity offset was added. 
        \label{fig:hires-singlechip}}
    \end{figure}

    The HIRES spectrograph has had a number of upgrades throughout its life on the Keck telescope.
    Two significant upgrades were the installation of its image rotator at the end of 1996, and the single-chip CCD that was upgraded to a mosaic of three CCDs in mid-2004.
    In the Keck sample of \citet{Murphy:2003em,Murphy:2004kl}, 77 of the 140 absorption system spectra were taken in the era before the image rotator was installed, while all of the spectra were taken during the ``single-chip'' era of HIRES.
    There were a number of nights of available archival data of solar twins and asteroids over the past 15 years.
    An example of the pre-image rotator supercalibration is shown in \Fref{fig:hires-singlechip}.
    The slope of the best-fit line to the data is 740 \msperthousand.
    While it is clear that there is a large, significant slope, it is also clear that a more elaborate model of the shape of the distortion could be fitted.
    However, we will fit a single straight line for the sake of simplicity. 

\begin{figure}
        \includegraphics[width=1.0\columnwidth]{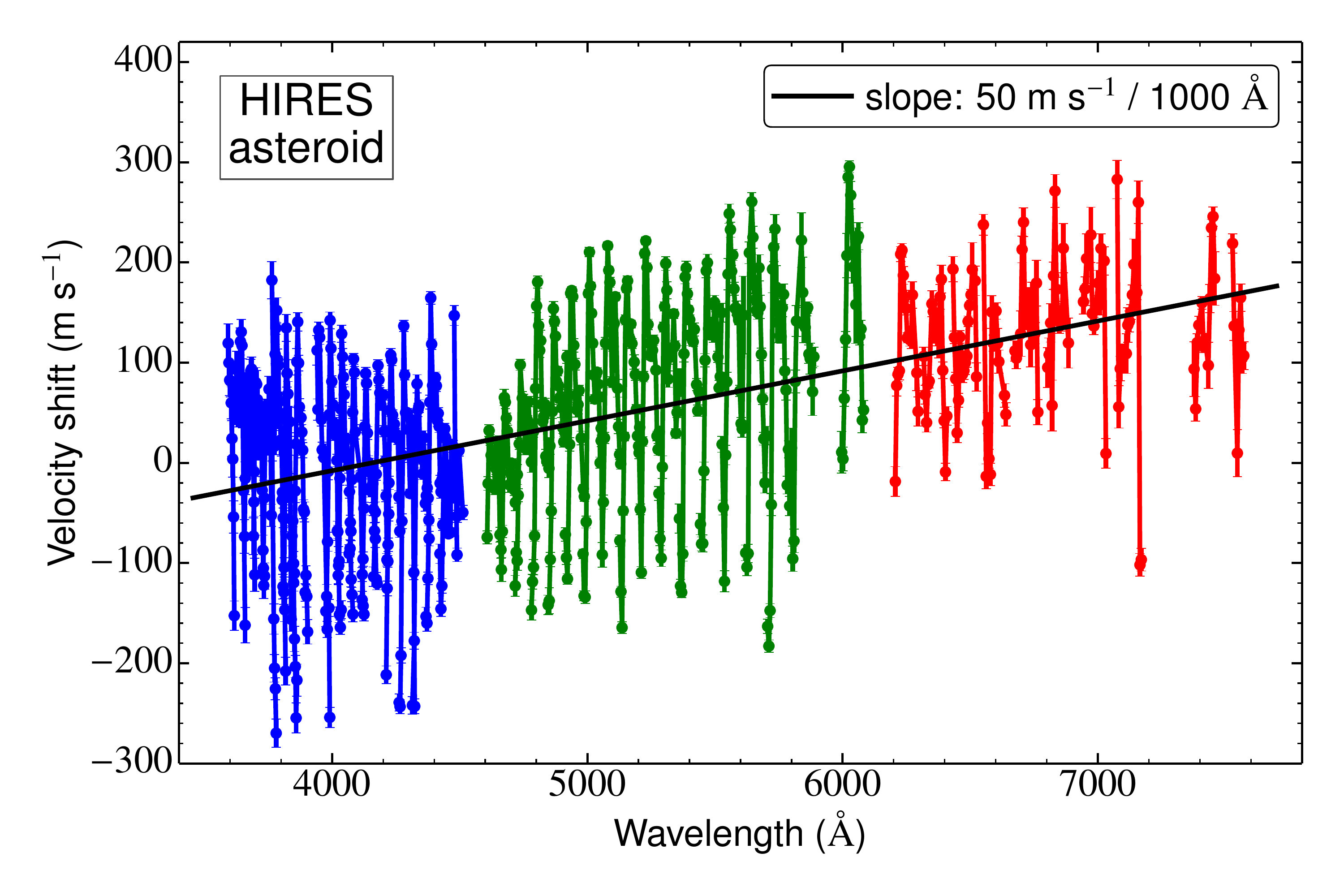}\vspace{-1.5em}
        \caption{The supercalibration result for an exposure of Ceres (asteroid) taken with HIRES on October 21, 2011.
        The exposure was taken with the image rotator in use during the 3-chip era of HIRES, and the resulting supercalibration velocity shift is plotted after an arbitrary constant velocity offset was added. 
        At wavelengths greater than 6000\,\AA, sky lines begin to dominate the spectrum.
        We mask these sky lines, which has the effect of producing gaps in the spectral coverage of the supercalibration results.
        \label{fig:hires-threechip}}
    \end{figure}

    During the 3-chip era, the image rotator was always used and the wavelength coverage was significantly improved.
    \Fref{fig:hires-threechip} shows the relatively small 50\,\msperthousand slope across the 3 CCD chips in an asteroid exposure taken on 21 Oct.~2011.
    The intra-order distortions are still evident as well; we do not observe any clear changes to their shape or amplitude with the transition to the 3-chip era. 

\begin{figure}
        \includegraphics[width=1.0\columnwidth]{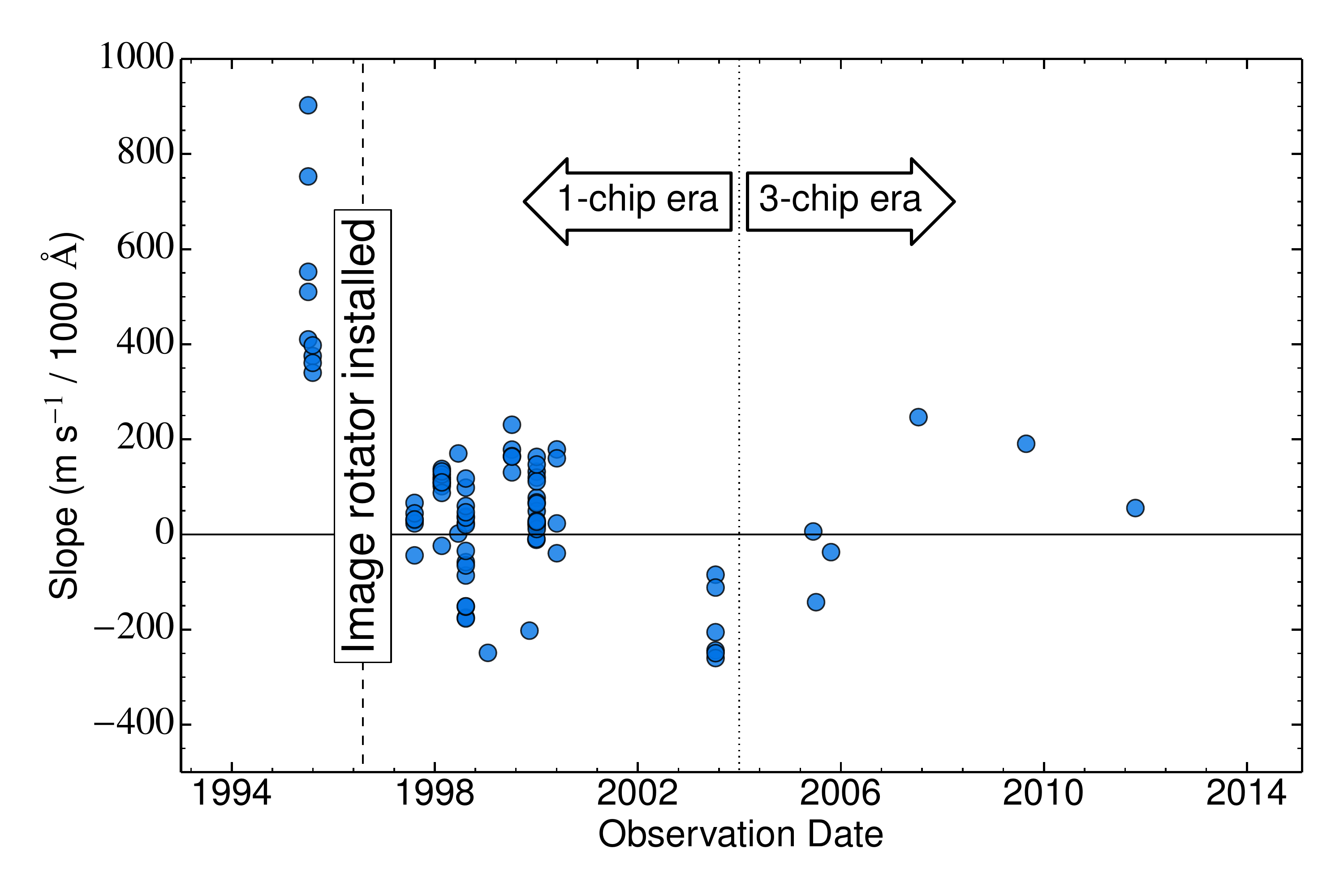}\vspace{-1.5em}
        \caption{The best-fit supercalibration slope for all archival HIRES solar twin and asteroid exposures we could identify with spectrograph setups similar to those used for the quasar spectra used by \citet{Murphy:2004kl}. 
        The vertical lines denote two upgrades to HIRES: the installation of the image rotator in late 1996 and the 3-CCD chip mosaic installation in 2004.
        All of the HIRES measurements used in the \citet{Murphy:2004kl}/\citet{2012MNRAS.422.3370K} studies were made prior to 2004.}
        \label{fig:hires-slope-history}
    \end{figure}

    We find a larger range in velocity distortion slopes in the archival data of solar spectra taken with HIRES compared with those of UVES.
    Each of the solar twin or asteroid exposures we could identify in the HIRES archive was supercalibrated against the solar spectrum as described in \Sref{sub:methods_solar}, and a best-fit linear slope was fit to the final supercalibration distortions. 
    The history of this linear slope over the past 15 years is plotted in \Fref{fig:hires-slope-history}.
    Generally, after the image rotator was installed, it appears that the long-range velocity slopes settle into a consistent range of $\sim\pm 200$\,\ms\ with no clear difference between the single-chip and 3-chip eras.
    However, while we could only identify two nights with useable data during the pre-image rotator era, it is clear that there are larger velocity distortions in those spectra.
    It is not clear at all from these pre-rotator data how well the slope we find might generalize to the rest of this era.
    Nevertheless, we note that, as observations were not made with the slit oriented with its spatial direction projected perpendicular to the physical horizon during this era, differential atmospheric refraction will have caused different wavelengths to fall at different positions across the slit. This would have led to velocity distortions \citep[e.g.][]{Murphy:2001uh,Murphy:2003em} and may be a contributing factor to the increased slopes observed in \Fref{fig:hires-slope-history} before 1996.

    \subsection{Quasar slit position effects} 
    \label{sub:results_sliteffects}

    There are fundamental differences between a ThAr and quasar exposure. 
    First, the ThAr calibration lamp light fills the slit during the calibration exposure, whereas the quasar presents a seeing disc at the slit.
    Second, the ThAr light is directed to the slit via a fold mirror, and any misalignment with respect to the quasar light path could lead to differential vignetting by the spectrograph optics.
    Finally, the ThAr light does not pass through the telescope optics, and any vignetting effects, for example, from the support structure of the secondary mirror, will not be present in the ThAr exposure \citep[e.g.][]{1995PASP..107..966V}.
    
    In addition to these inherent differences, the practical goal of keeping a quasar centred in the slit during long exposures is often difficult to maintain. 
    The quasar often drifts within -- both along and across -- the slit.
    Drifts across the slit are particularly problematic because they translate to velocity shifts in the spectral lines and possibly higher-order effects leading to velocity distortions.
    For example, the effective instrument profile will change as one side of the quasar seeing disc is vignetted by the slit.
    If that effect has a wavelength-dependent component (e.g.~seeing dependence on wavelength), a velocity distortion could be induced.
    Also, as discussed above (\Sref{ssub:ssub:results_longrange_hires}), if observations are not made at the parallactic angle, differential atmospheric refraction may cause long-range velocity distortions. However, as discussed in \citet{Murphy:2001uh,Murphy:2003em}, this is only relevant for HIRES quasar spectra taken before the image rotator was installed in 1996; all other spectra were observed with the slit held at the appropriate angle.

    Here we attempt to quantify miscalibrations due to such quasar position effects by taking several exposures of bright standard stars with the iodine cell in place and deliberately placing the quasar at different positions across the slit.

    \subsubsection{VLT--UVES} 
    \label{ssub:results_sliteffects_uves}

    Two fast-rotating, bright stars, HR9087 and HR1996, were observed in 2009 with VLT--UVES with a 0.7'' slit and ``attached'' ThAr exposures.
    Three exposures of each star were recorded with the iodine cell in place in the following way: the star was displaced from the slit centre by about a third of the slit width and an exposure was taken.
    The displacement was made in the spectral direction alone, i.e. it was not moved along the slit, but only across the (short) width of the slit. 
    The star was then placed in the centre of the slit and a second exposure was taken, followed by a third exposure with the star displaced by about a third of the slit width in the opposite direction to the first exposure.
    Finally, the iodine cell was removed from the light path and an attached ThAr was taken. 
    This entire procedure was done within a single observation block (OB), so there was a single attached ThAr exposure for the three corresponding star exposures.

\begin{figure}
        \includegraphics[width=1.0\columnwidth]{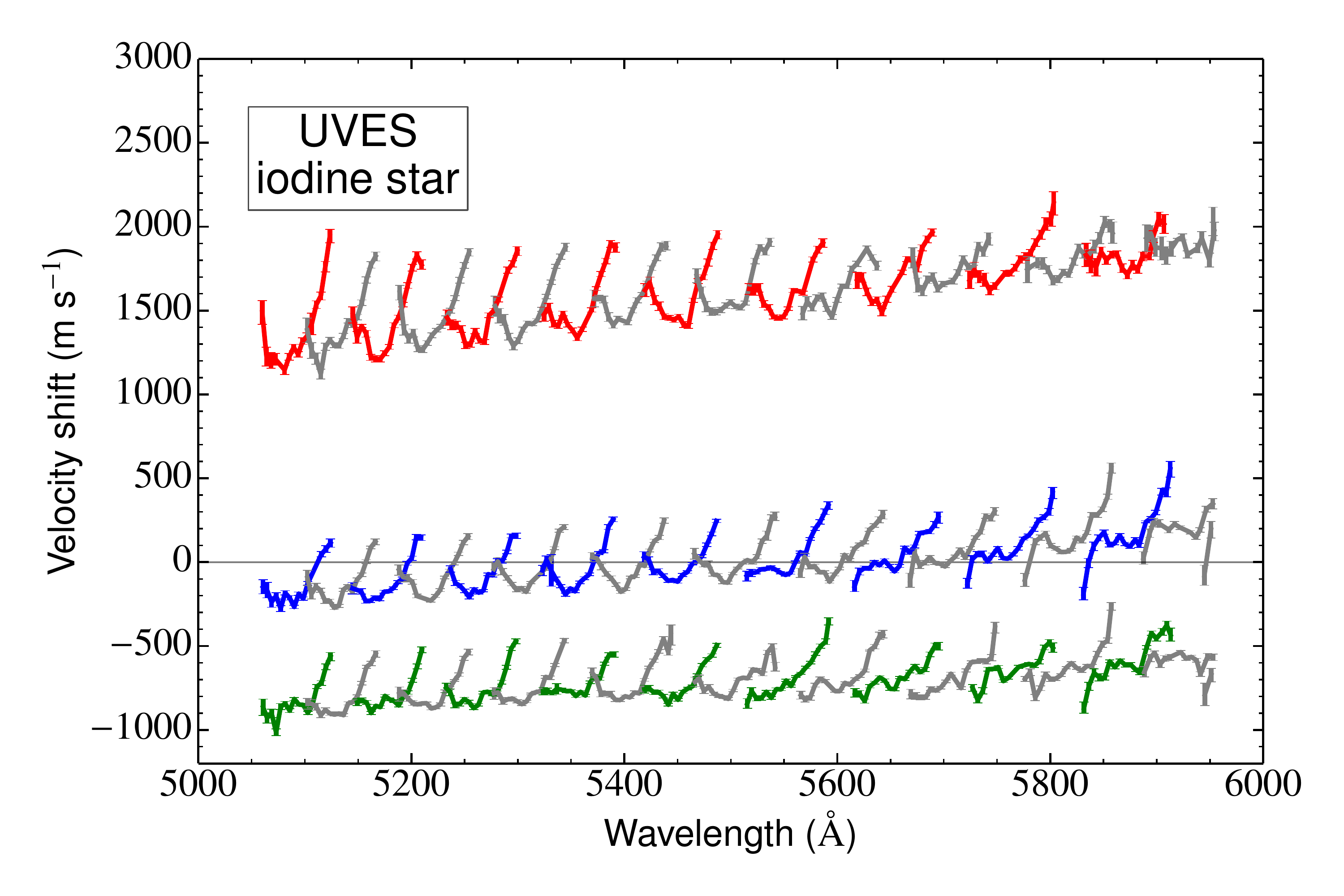}\vspace{-1.0em}
        \includegraphics[width=1.0\columnwidth]{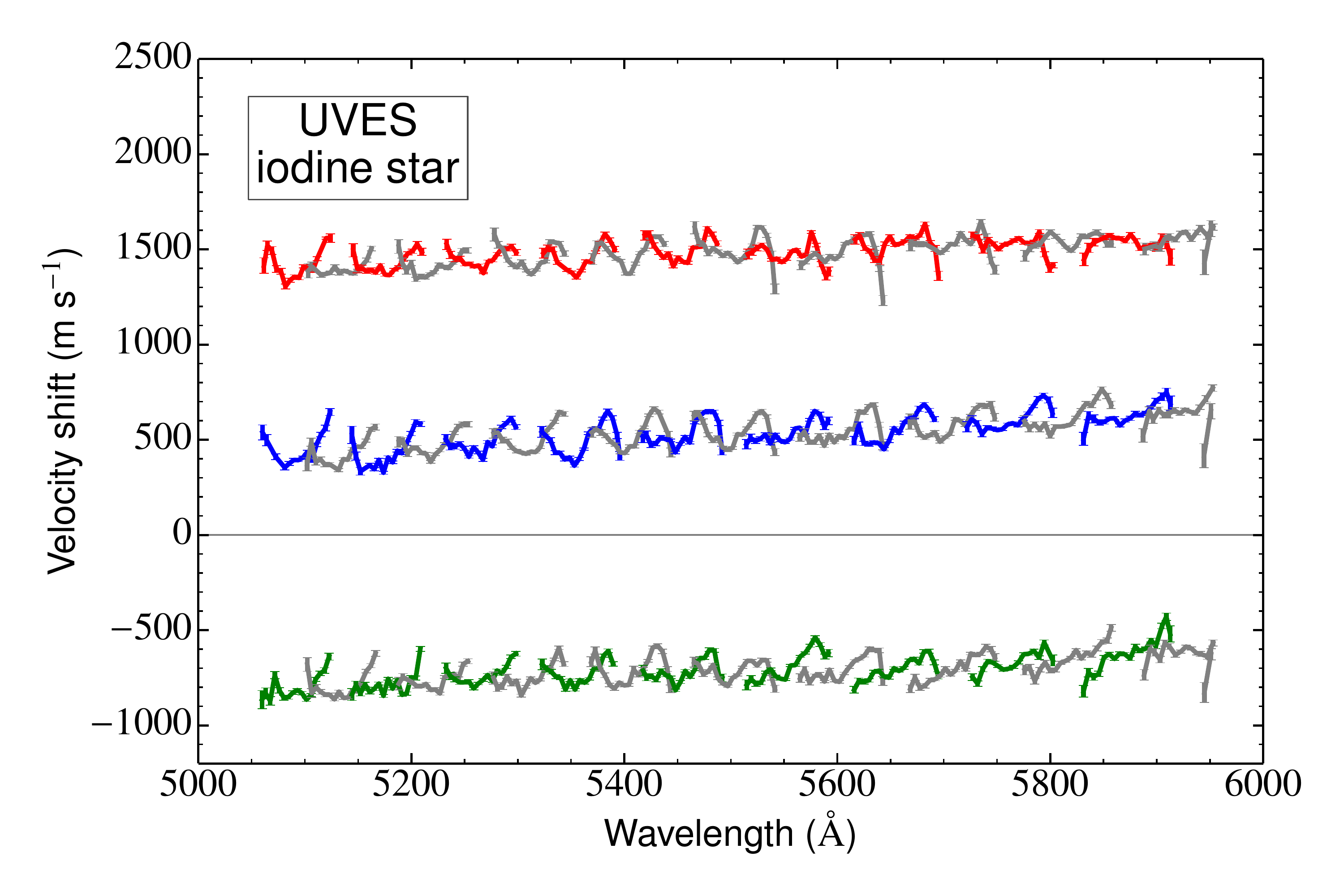}\vspace{-1.5em}
        \caption{Iodine cell supercalibration measurements of  stars observed on UVES.
        The stars were deliberately displaced across the slit in three positions while the iodine cell was in the light path.
        Upper plot shows the three supercalibration exposures of HR9087, and lower plot shows the three HR1996 exposures.
        Within each exposure, we plot alternating colors to distinguish adjacent orders, and no velocity shift is applied.
        \label{fig:vlt-slit-test}}
    \end{figure}

    The iodine cell supercalibration was applied to each of the star exposures (each calibrated with the single ThAr exposure) and the results are shown in \Fref{fig:vlt-slit-test}.
    The shifts in the star slit position appear to translate only to a constant velocity shift, as expected, and by the expected amount (i.e.~$\sim$2 \kms).
    The intra-order and long-range velocity distortions are also clearly apparent.
    However, neither of these appear to change in a deterministic way with the star slit position.
    One final note is that these exposures were taken within a single night, and yet the long-range supercalibration slope is $\approx$400\,\msperthousand in each of the HR9087 (upper plot) spectra, while $\approx$100\,\msperthousand in the HR1996 (lower plot) spectra. 

    \subsubsection{Keck--HIRES} 
    \label{ssub:results_sliteffects_hires}
    
\begin{figure}
        \includegraphics[width=1.0\columnwidth]{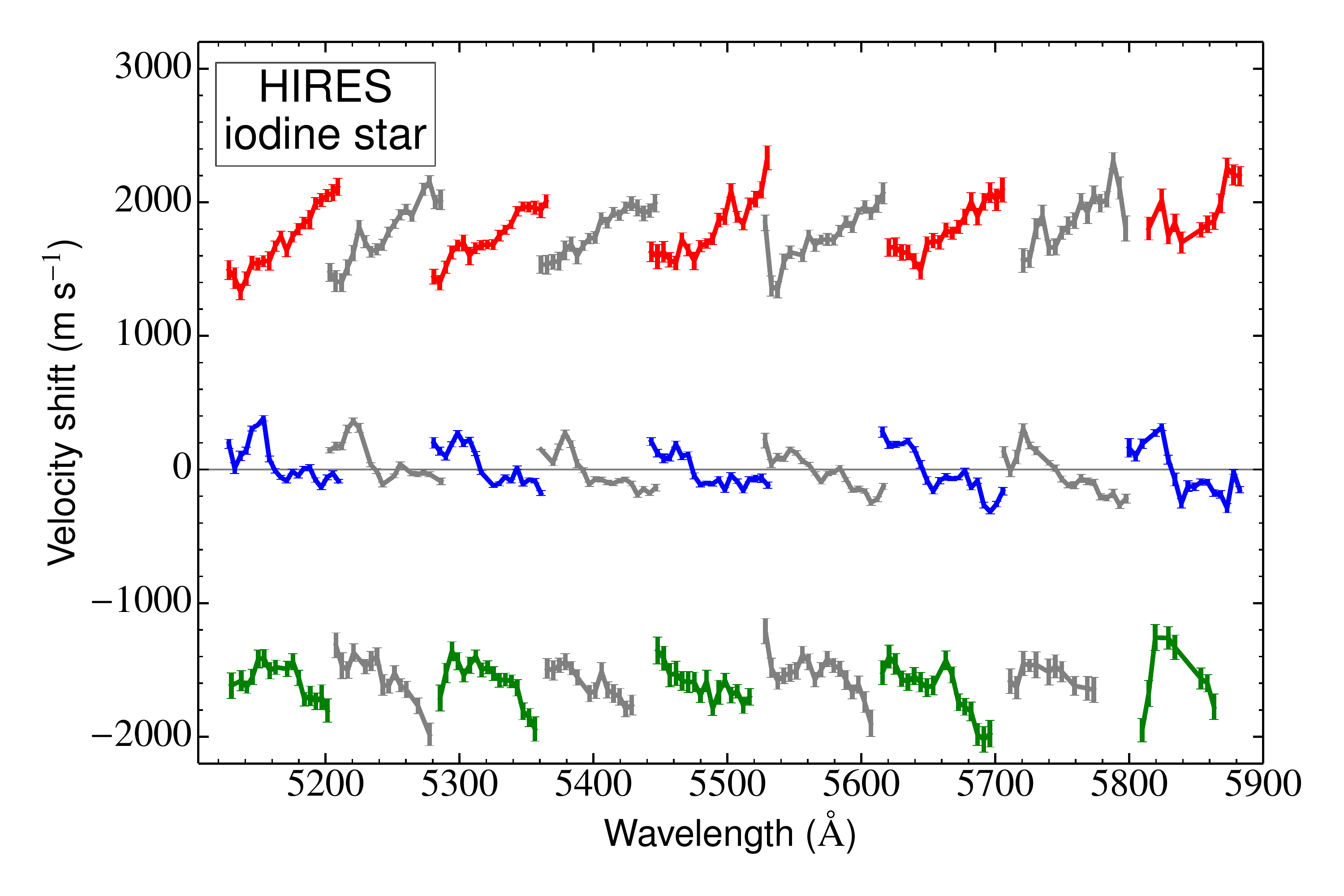}\vspace{-1.5em}
        \caption{Three iodine cell supercalibration measurements of star Hiltner 600 observed on HIRES while deliberately displacing the star to three positions across the slit.
        Within each exposure, we plot alternating colors to distinguish adjacent orders, and no velocity shift is applied.
        \label{fig:hires-slit-test1}}
    \end{figure}

\begin{figure}
        \includegraphics[width=1.0\columnwidth]{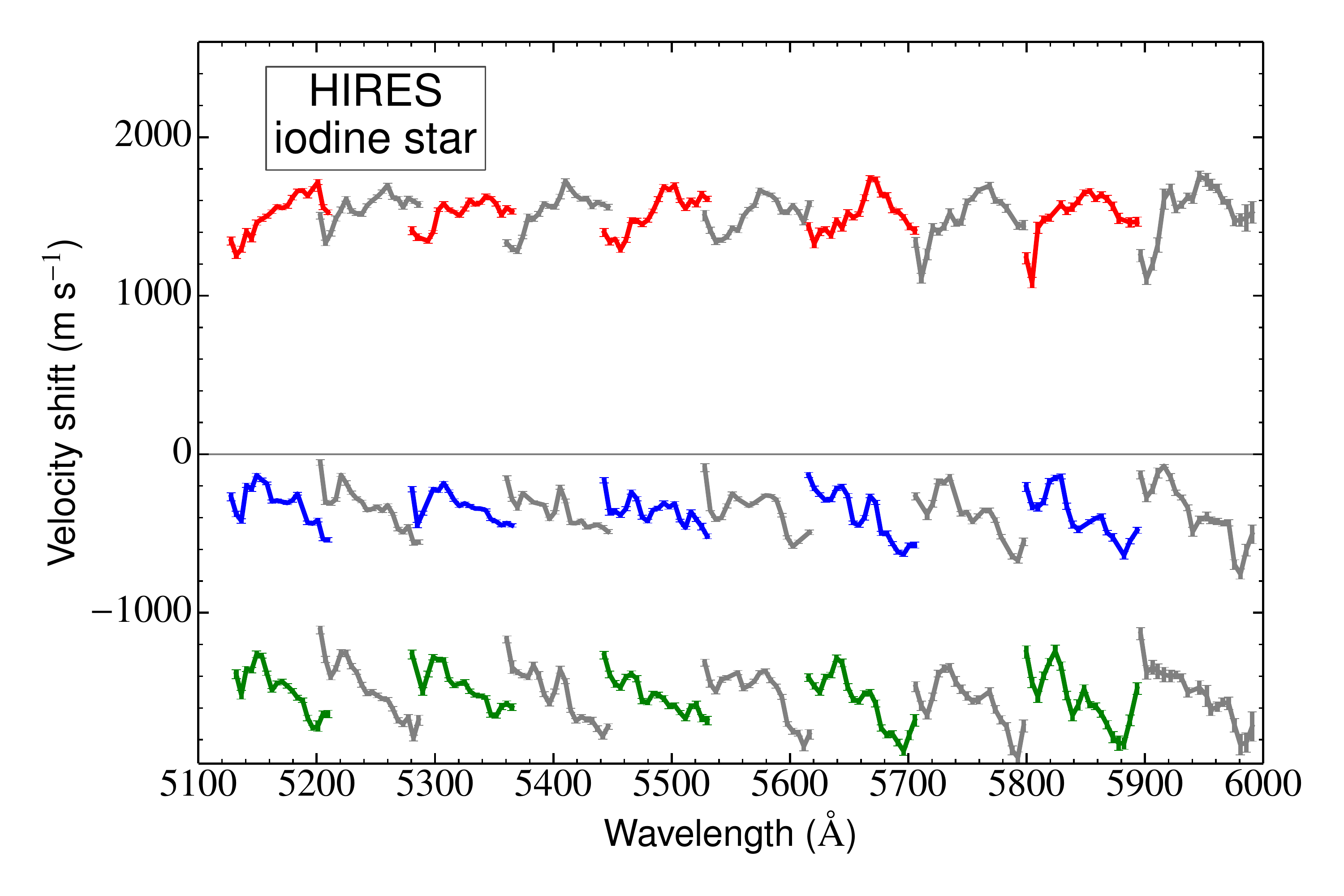}\vspace{-1.0em}
        \includegraphics[width=1.0\columnwidth]{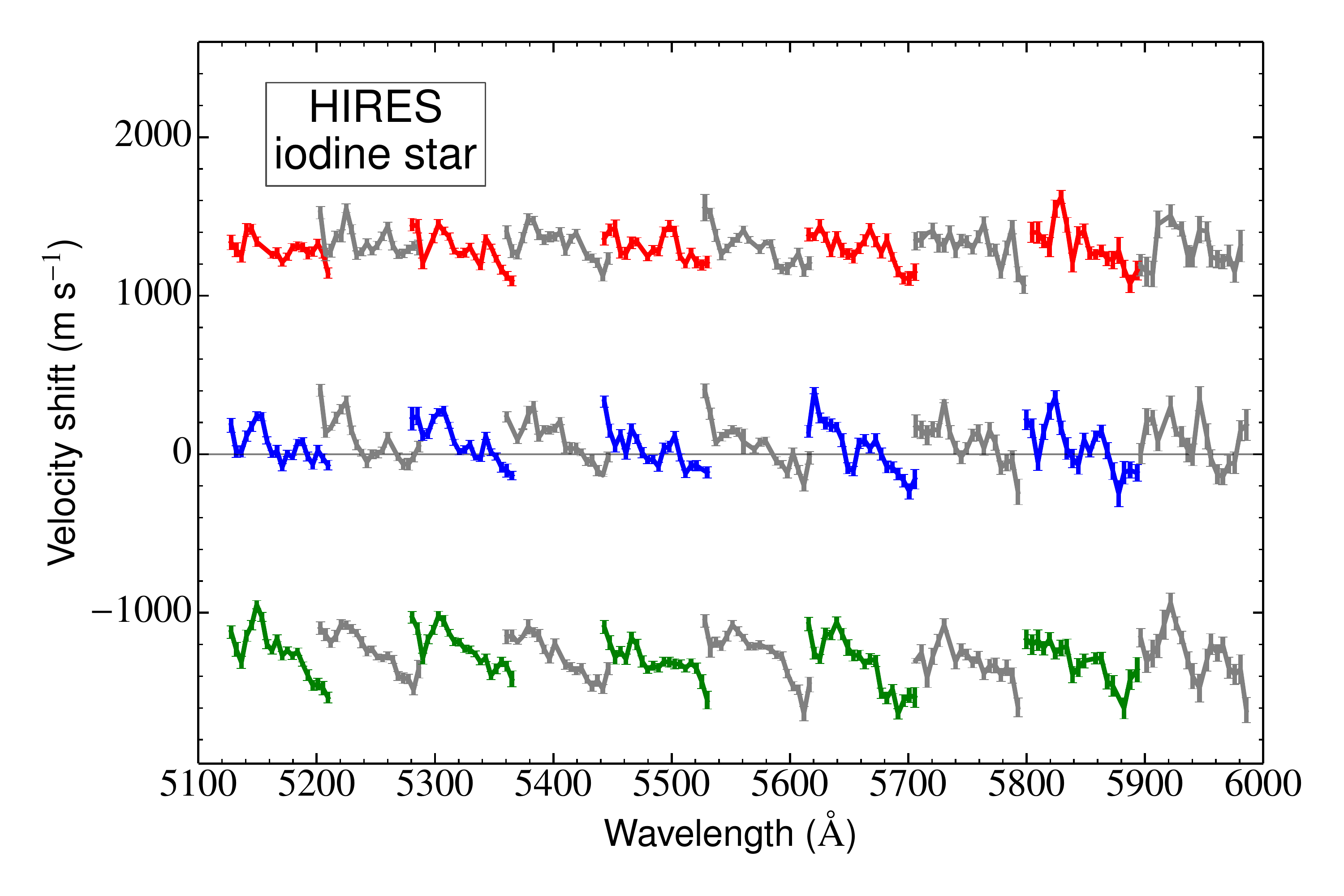}\vspace{-1.5em}
        \caption{Three iodine cell supercalibration measurements of stars HR9087 (upper) and GD 71 (lower) observed on HIRES while deliberately displacing the star to three positions across in the slit.
        Within each exposure, we plot alternating colors to distinguish adjacent orders, and no velocity shift is applied.
        \label{fig:hires-slit-test2}}
    \end{figure}

    In contrast to UVES, the gratings of the HIRES spectrograph are not automatically reset after each OB (indeed there is no OB structure or concept when using HIRES).
    Instead, observers leave the spectrograph's gratings in place and take a ThAr before the telescope slews to a new object. 
    This allows for a more simple procedure when conducting the same slit position experiment and this was undertaken with HIRES in November 2009 using a 0.''861 slit, with the star Hiltner 600, with the results shown in \Fref{fig:hires-slit-test1}.
    A second test was conducted a month later using the stars HR9087 and GD 71, whose results are shown in \Fref{fig:hires-slit-test2}.
    In HIRES, it appears that the intra-order distortions might change in shape systematically with the placement of the star across the slit (cf. UVES).
    However, this does not appear to have any effect on the long-range distortion slope which seems nearly flat for these exposures. 

    \subsection{Stability of the VLT--UVES wavelength scale} 
    \label{sub:ssub:results_uvesstability}
    
    While the supercalibration method in \Sref{sec:methods} was used to derive absolute velocity distortions, the \emph{relative} stability of the ThAr wavelength solutions can be tested in a more direct manner: by comparing ThAr wavelength \emph{solutions} from exposures taken over a short timespan to each other.
    As shown in \Sref{ssub:results_longrange_uves}, for most UVES exposures there are clear long-range velocity distortions.
    Here we investigate whether there are drifts in the spectrograph over $\sim$20 minute time-scales which change these distortions.

\begin{figure}
        \includegraphics[width=1.0\columnwidth]{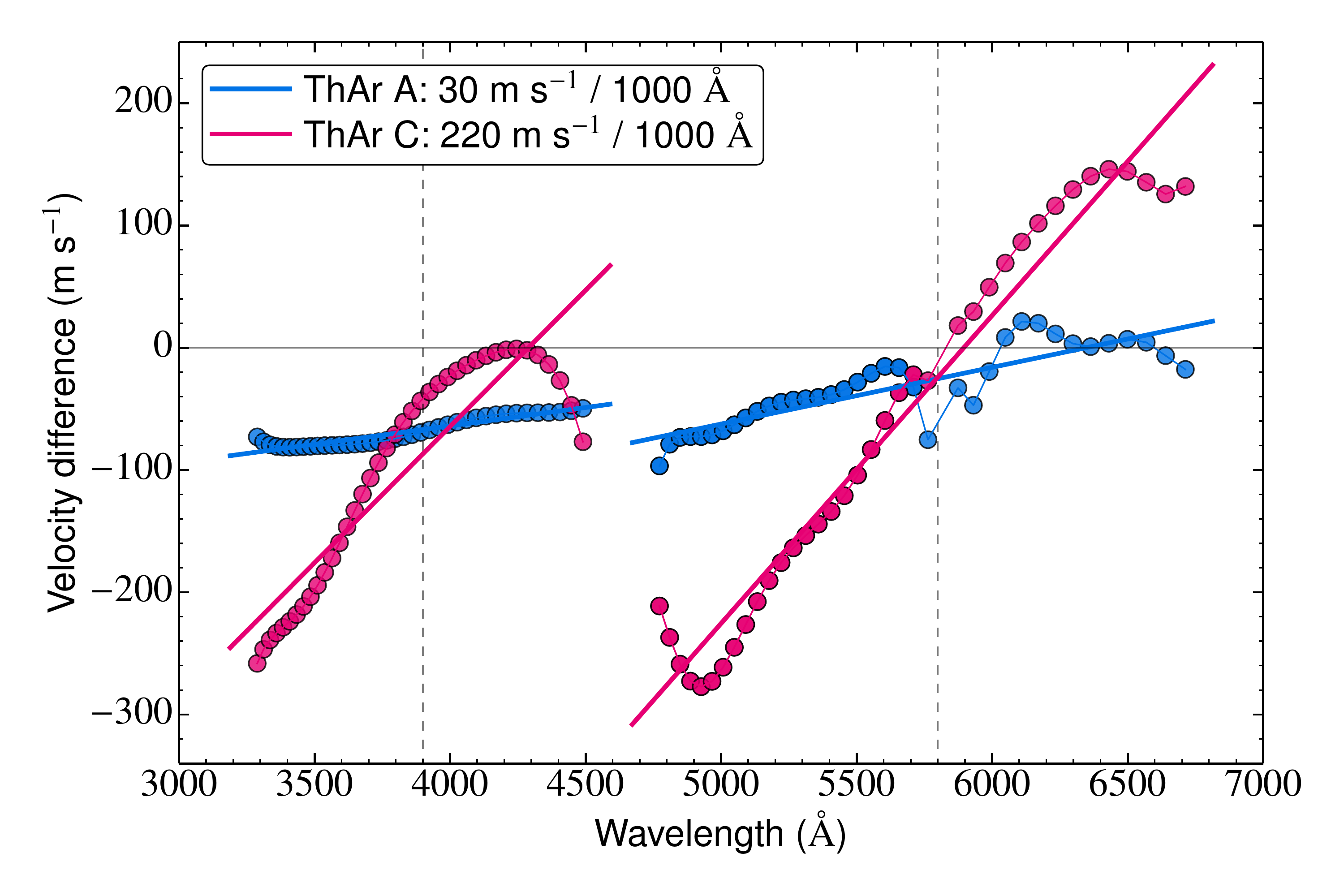}\vspace{-1.5em}
        \caption{
        Comparison of wavelength solutions from ``unattached'' ThAr exposures (points) to that derived from an ``attached'' one (zero velocity shift at all wavelengths).
        We compare two ``unattached'' ThAr wavelength solutions directly to the ``attached'' ThAr wavelength solution.
        The two ``unattached'' ThAr exposures were taken roughly 10 minutes before and 10 minutes after the ``attached'' ThAr exposure.
        Each point represents the average difference of the central 80 pixels of each order in the wavelength solution between the ``attached'' and the ``unattached'' ThAr wavelength solutions.
        The y-axis shows the shift plotted in velocity space between the two solutions and the x-axis shows the location of the order in wavelength space.
        The vertical dashed lines denote the central wavelength setting of each arm: 3900\,\AA\ for the blue arm, and 5800\,\AA\ for the red arm.}
        \label{fig:solution-subtraction}
    \end{figure}

    For three science--ThAr exposure pairs (call them A, B, and C) with ``attached'' ThAr calibration exposures, taken about ten minutes apart, we reduced science exposure B with each ThAr exposure.
    In other words, the ThAr exposures A and C were effectively ``unattached'' with respect to science exposure B, while ThAr exposure B was ``attached'' to science exposure B.
    We compare the ``attached'' wavelength solutions with the ``unattached'' ones in \Fref{fig:solution-subtraction} by simply subtracting ThAr exposure B's wavelength solution from those of ThAr exposures A and C. Figure \ref{fig:solution-subtraction} shows that there is clearly an instability in the ThAr wavelength solution over $\approx$20 minute timescales that can produce a relative long-range velocity distortion of several hundred \ms over 1000 \AA.
    Further, the relative slope appears to be shared between the two arms, while the relative offset between the central wavelength of each arm appears to be relatively aligned.
    We stress that this is the relative slope because this is a slope only in comparison to the ``attached'' ThAr wavelength solution. 
    This behavior is remarkably similar to that found in the absolute velocity distortions of \Sref{ssub:results_longrange_uves}, which suggests that the cause of the long-range distortions is likely a physical effect within the spectrograph.

\begin{figure}
        \includegraphics[width=1.0\columnwidth]{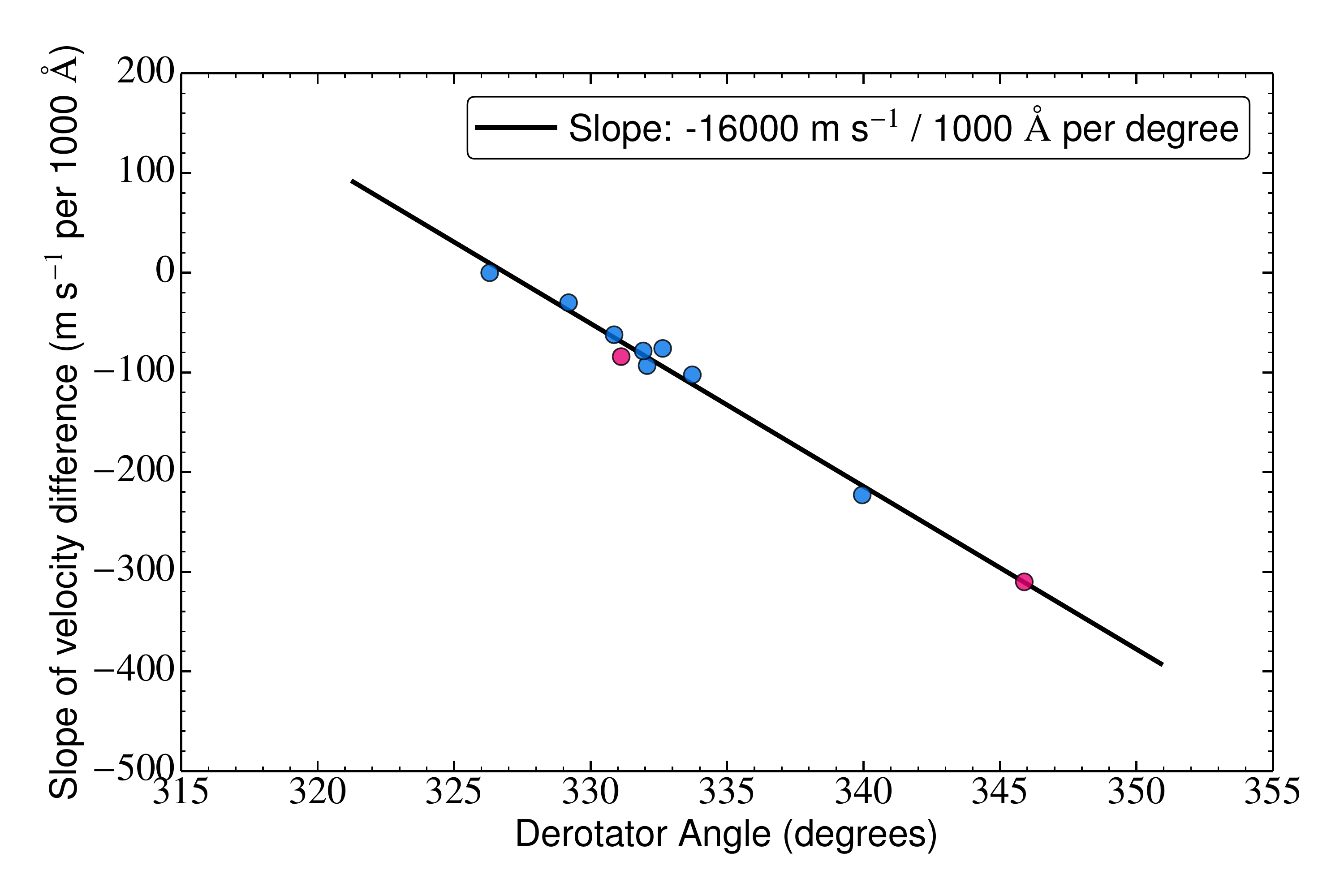}\vspace{-1.5em}
        \caption{
        Slope of the difference between ThAr wavelength solutions taken over a single night (blue points) and the preceding night (red points) versus the derotator angle recorded in the UVES exposures' headers. 
        The line shows the best-fit linear slope to these points, indicating an extremely tight correlation between the two quantities. 
        We do not suggest extrapolating this trend because, in ongoing tests to be reported in future, the relationship appears to be sinusoidal when considering a broader derotator angle range.
        }
        \label{fig:derotator-relationship}
    \end{figure}

    We have identified the apparent cause of this effect by extending the same analysis to other ThAr exposures taken throughout that night and two taken the previous night.
    \Fref{fig:derotator-relationship} shows the relationship between the slope of the best-fit linear velocity distortion in the blue CCD against the derotator angle.
    The two red points are the ThAr exposures taken the previous night, and they also appear to align well with the overall slope of the trend.
    The tightness of the correlation suggests that these relative distortions arise due to the change in derotator angle with only a small possible contribution from other effects like mechanical and/or temperature drifts in the spectrograph. 
    The exposures cover only a small derotator angle range and, in ongoing tests to be reported in a future paper, it appears likely that the trend does not continue linearly but rather cycles sinusoidally.
    It is surprising that the image rotator would itself induce a long-range velocity distortion in the ThAr spectrum, and it remains unclear whether, and in what manner, the effect would potentially propagate to the quasar spectrum as well. 
    Again, this is only the intrinsic relative instability in the ThAr wavelength solution, and the absolute distortions in the wavelength scale of the quasar spectrum cannot be inferred from it except by a comparison method like supercalibration.

    Finally, we undertook a similar analysis with Keck--HIRES, but on that spectrograph the ThAr light does not pass through the image rotator (unlike on UVES), and no relationship between rotator angle and any relative slope was found.

    \subsection{Summary and hypothesis} 
    \label{sub:results_summary}

    The distortion results of the supercalibration method can be summarized as follows. 
    In both UVES and HIRES, tests of the effect of deliberate slit position offsets shows that the position of an astronomical object across the slit will give rise to a constant velocity shift to the spectrum, while a mis-centreing does not appear to induce long-range velocity slopes.
    These constant velocity shifts have the effect of smearing the spectrum and degrading the overall signal in coadded spectra, but will not add to a relative velocity shift between lines, except in cases where different wavelength coverages are coadded \citep{2013ApJ...778..173E}.
    Also, the guiding will move the object in the slit, especially during long exposures typical of quasar observations.
    The slit position effects will therefore be present in most quasar exposures but they will not greatly affect varying-$\alpha$ analyses. 

    Much more important are the intra-order and long-range velocity distortions, both of which are ubiquitous in both spectrographs.
    The long-range distortions are particularly concerning for varying-$\alpha$ analyses.
    It appears that in UVES there tends to be a ``lightning bolt'' distortion in the wavelength scale across the blue and red arms of the spectrograph.
    In HIRES it appears that, in most cases, the same distortion applies across all 3 chips in the single arm, albeit with some flattening in the bluest $\sim$3rd of the wavelength coverage. 
    The long-range distortions do not appear to change with the slit position of the object in either spectrograph, whereas the intra-order distortion shapes may change due to this effect in HIRES.

    With these facts in mind, we may establish a working hypothesis for the cause(s) of the long-range distortions, though we do not claim to provide an ultimate explanation here.
    We propose that the effective instrument profile (IP) changes across the spectrographs' focal planes in both the spectral and spatial directions, according to some vignetting effect within the spectrographs.
    It has been shown previously that the HIRES IP varies across an order in iodine star studies \citep{1995PASP..107..966V,1996PASP..108..500B}; we assume the UVES IP must vary by a similar magnitude.
    If the IP variation is significantly different for the ThAr calibration exposure, it may lead to the calibration distortions we observe.
    Because the long-range distortions are not noticeably sensitive to astronomical slit position, if this hypothesis is correct then the vignetting effect must be most prominent and variable in the ThAr calibration exposures.
    This possibility is supported by \Fref{fig:derotator-relationship}, which shows large changes in the distortion slope when different ThAr exposures, taken at different (de)rotator angles, are compared.
    Differential changes in the IP across the focal plane, between the ThAr and astronomical object exposures, will be degenerate (to first order) with apparent relative velocity shifts between transitions at different wavelengths.
    That is, long-range and intra-order velocity distortions will result, and cause systematic effects in \daa measurements.

\section{Implications for previous varying $\alpha$ studies} 
\label{sec:implications}
    
    The long-range velocity distortions identified in \Sref{sec:results} have direct implications for measuring \daa with UVES and HIRES.
    In this section, we analyze the potential impact of these distortions on previous studies by applying a simple model of the measured supercalibration distortions to simulated data.
    We then fit for \daa on this distorted simulated data in an attempt to quantify the effect of the velocity distortions.
    The absorption systems that we consider come from the VLT--UVES measurements of \citet{2011PhRvL.107s1101W} and \citet{2012MNRAS.422.3370K} and the Keck--HIRES measurements of \citet{Murphy:2003em,Murphy:2004kl} and \citet{2012MNRAS.422.3370K}.
    
    The simulated data for both UVES and HIRES were generated with the same process, and we applied the distortion model of each spectrograph to its respective absorption system sample. 
    We used \textsc{rdgen} 10.0\footnote{Both {\sc vpfit} and {\sc rdgen} are available at \href{http://www.ast.cam.ac.uk/~rfc/vpfit.html}{\url{http://www.ast.cam.ac.uk/~rfc/vpfit.html}}} to simulate each absorption system spectrum with a pixel-size of 1.3 \kms, a \SNR of 2000\,pix$^{-1}$, and a Gaussian instrument profile (IP) FWHM of 5.0 \kms.
    We placed each simulated absorption system at the corresponding measured redshift of each system in the HIRES or UVES sample. 
    In contrast to the measured systems, these simulated systems were created with a single velocity component with Doppler parameter $b = 2.5$ \kms.
    We use a single component velocity structure in our simulated data instead of the multi-component models fitted to the real data in the original analysis for the sake of simplicity and to focus attention on the effect of the distortions on \daa rather than possible effects of a complicated velocity structure. 
    Finally, the column densities, $N$, for the different ionic species were assigned in relative proportion similar to the simple model in \citet{Murphy:2014hu} based on the solar abundance pattern for the singly ionized species (which will dominate in damped Lyman-$\alpha$ systems)\footnote{Specifically, the \logN assigned to each species was (in units of cm$^{-2}$): \ion{Mg}{i} 11.28; 
     \ion{Mg}{ii}  13.08; 
     \ion{Al}{ii}  11.98; 
     \ion{Al}{iii} 11.38; 
     \ion{Si}{ii}  13.06; 
     \ion{Cr}{ii}  11.18; 
     \ion{Fe}{ii}  13.00; 
     \ion{Mn}{ii}  11.03; 
     \ion{Ni}{ii}  11.75; 
     \ion{Ti}{ii}  11.46; and 
     \ion{Zn}{ii}  11.18.}.
    To give an explicit example: consider an absorption system in the VLT--UVES analysis of \citet{2012MNRAS.422.3370K} at a redshift $z=2.7686$ in which the following transitions were used to measure \daa: \ion{Fe}{ii} 1608, \ion{Si}{ii} 1526, and \ion{Si}{ii} 1808.
    Correspondingly, in our simulation we simulated a spectrum with a single velocity-component absorption system at the same redshift, $z=2.7686$, with $b=2.5$ \kms\ and the column densities for those transitions of $\log[N($Fe{\sc \,ii}$)/$cm$^{-2}] = 13.00$ and $\log[N($Si{\sc \,ii}$)/$cm$^{-2}] = 13.06$.

    We then distort this simulated spectrum by applying a simple model of the long-range velocity shift to the wavelength scale using \Eref{eq:vshift}. 
    We fit for \daa on the final distorted spectrum, and since we simulated the spectrum with $\daa=0$, any non-zero \daa must result directly from the applied distortion model (this is why the SNR used in the simulation is very high, 2000\,pix$^{-1}$; the statistical error on \daa will therefore be negligibly small).
    The process of fitting for \daa that we adopt is the same approach used in \citet{Murphy:2004kl} and \citet{2012MNRAS.422.3370K}:
    (i) use \textsc{vpfit} 10.0 to fit a single-component absorption feature [fixing $\daa=0$] to create an initial fit; and
    (ii) run this initial fit, this time allowing \daa to vary to give the final fit. 
    In all of these tests, the $b$, $z$, and $\alpha$ parameters are tied together across all species, while the \logN value for each species is allowed to vary. 
    In this way, we run a simulated experiment of an ensemble of observations for a given velocity distortion model.

    \subsection{VLT--UVES simulation} 
    \label{sub:implications_uves}

    We create the simple VLT distortion model by capturing the most striking characteristics of the VLT distortions: that each spectrograph arm tends to share a similar velocity distortion slope (both in sign and magnitude), and the central wavelengths of each arm tend to be aligned in velocity, i.e. there is no overall velocity shift between the arms.
    That is, there is just a single parameter required to describe the model -- the distortion slope of both arms. 
    To define this parameter, we use the supercalibration information we have for the era in which the quasar spectra used by \citet{2012MNRAS.422.3370K} were taken, i.e.~prior to 2009 (see \Fref{fig:uves-slope-history}). 
    For each night of observations in that era, we average the slopes for the supercalibration exposures of the bluer of the two chips in the red arm of UVES. 
    The slope used in our simple model is the median of these nightly averages prior to 2009, i.e.~117\,\msperthousand. 
    While this value typifies the slopes found, it should be noted that slopes as large as 600 \ms were measured for UVES with the supercalibration method (e.g.~\Fref{fig:uves-distortion}) and we have very sparse information from the archival search of supercalibration spectra.

\begin{figure}
        \includegraphics[width=1.0\columnwidth]{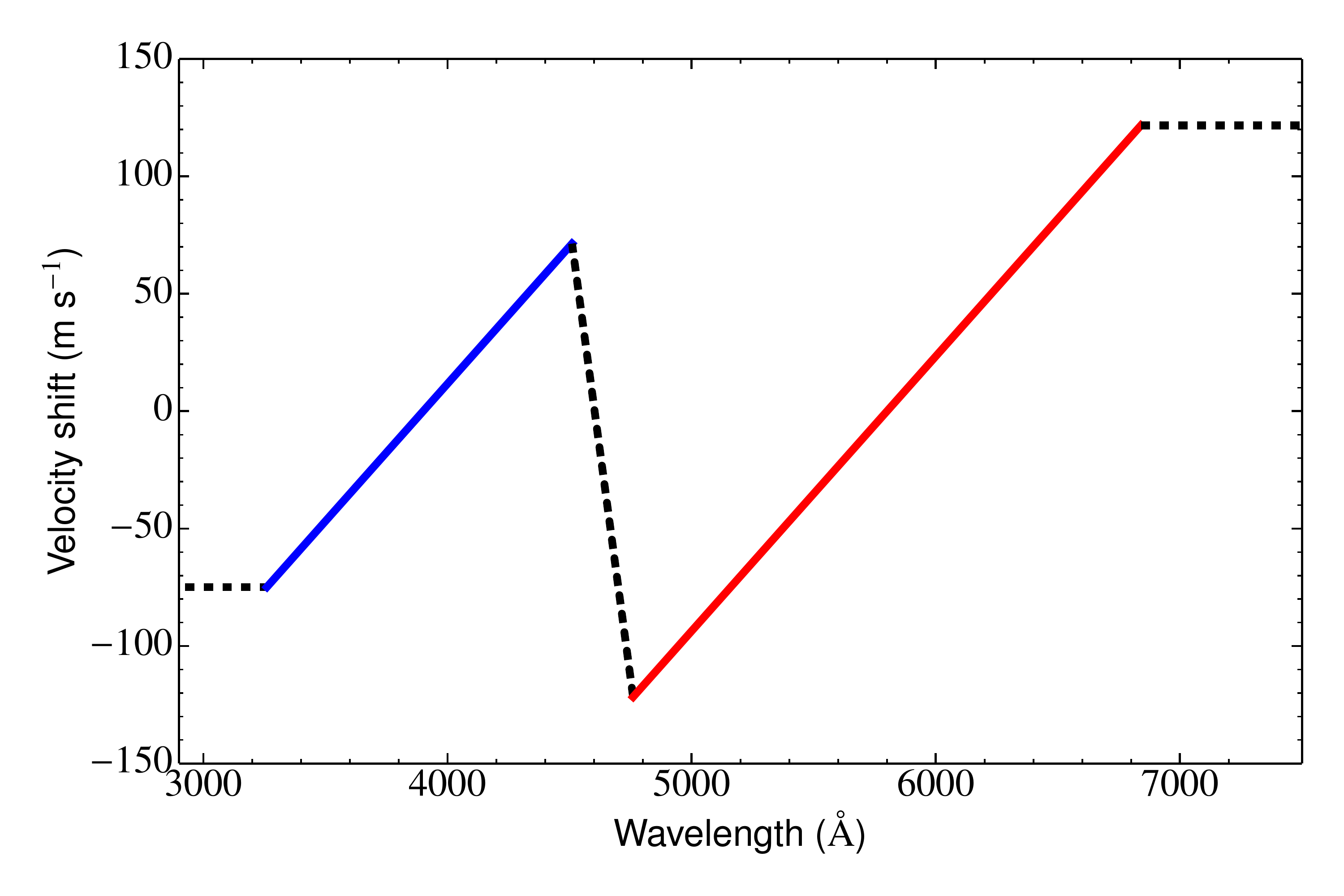}\vspace{-1.5em}
        \caption{The simple distortion function that we adopt to model the velocity distortions found in UVES.
        This is a five-piece linear model with a slope of 117\,\msperthousand in the blue and red arms of a typical 390/580-nm setting on UVES.
        Outside of that setting's wavelength range, we simply extend the offset with a constant value, while between the blue and red arm, a straight line connects the end of the blue arm model to the beginning of the red arm model. 
        The slope of 117\,\msperthousand derives from the median of the nightly average supercalibration slopes from the archival information prior to 2009.
        \label{fig:vlt-split-linear-model}}
    \end{figure}

    \Fref{fig:vlt-split-linear-model} shows the simple distortion model that we implemented.
    We model the distortion to reflect the wavelength coverage of the 390/580 setting.
    This setting corresponds to the setting used in the majority of measurements found in the VLT--UVES sample in \citet{2012MNRAS.422.3370K}.
    The full VLT--UVES sample uses wavelengths not covered by the 390/580 setting so, in other regions outside of the its wavelength coverage, we have simply extrapolated the distortion model with a constant velocity shift value at the extreme blue and red wavelengths.
    As the distortion is centred on the central wavelength of each arm's setting, extending the 580 setting to longer wavelengths would exaggerate the distortion unrealistically.
    Finally, in the region between the end of the blue arm and the beginning of the red arm we simply connect the end-points with a straight line to preserve continuity.
    This approach yields a model that we can easily adopt for all of the systems measured.
    
\begin{figure*}
        \includegraphics[width=0.75\textwidth]{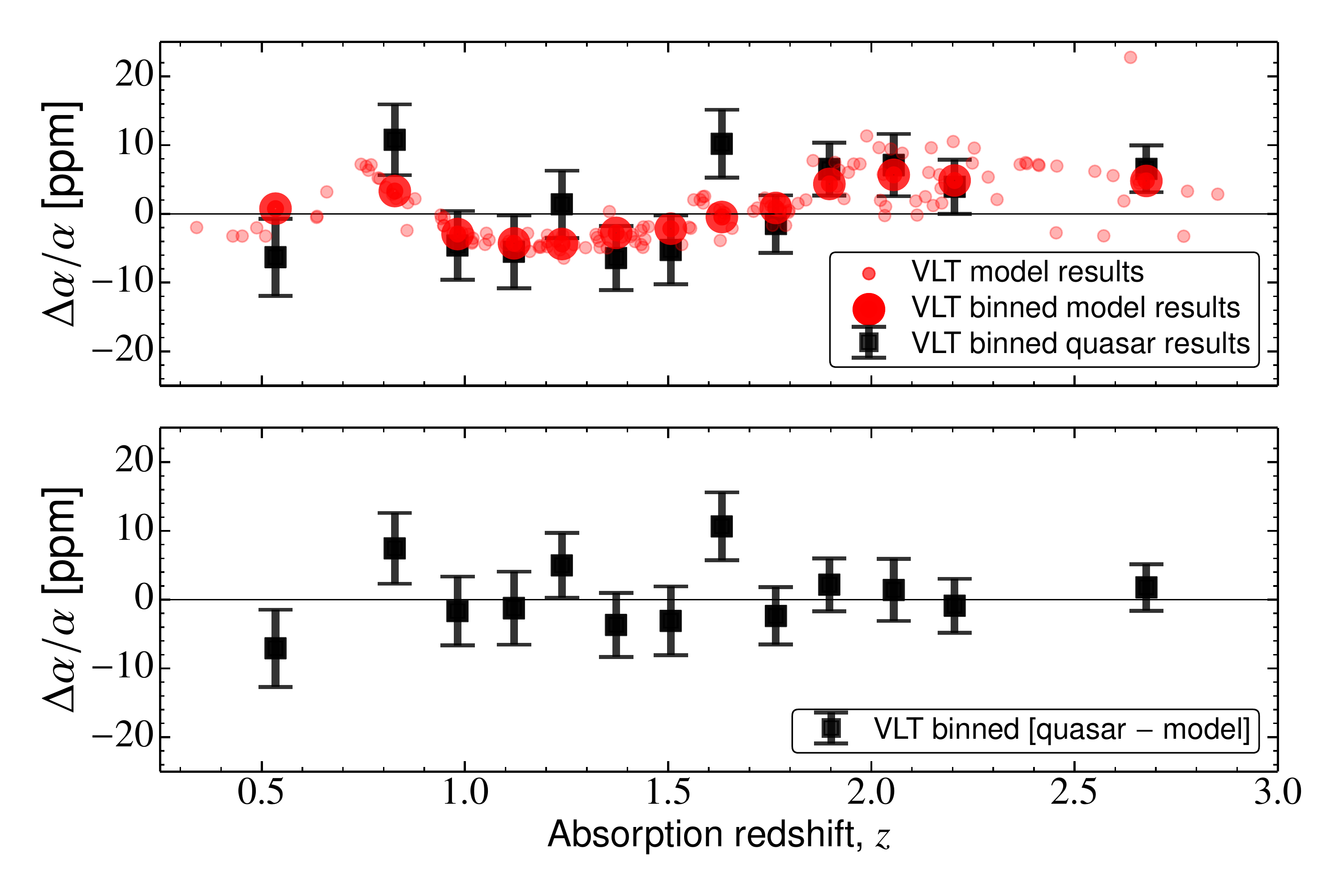}\vspace{-1.5em}
        \caption{
        The upper panel compares the binned \daa measurements of \citet{2012MNRAS.422.3370K} from VLT--UVES quasar spectra (black squares with 1-$\sigma$ errors) with the results of applying our simple distortion model (\Fref{fig:vlt-split-linear-model}) to simulated spectra of those UVES absorption systems (small red/grey filled circles).
        The large red/grey filled circles show the weighted mean of the simulated \daa values in the same bins as the quasar measurements and using the inverse-squares of the quasar 1-$\sigma$ uncertainties as weights.
        Each bin contains $\approx$12 of the 153 \daa values from both datasets.
        Although the model of the long-range distortion is very simple and will not reflect the real distortions in all exposures of all quasars, the simulated results reflect many of the characteristics of the real quasar measurements, e.g.~the overall reversal in sign from low to high redshifts, but with similar magnitudes and possibly a similar deviation at $z\approx0.8$.
        The lower panel shows the \daa values from the quasar spectra after subtracting the simulated \daa values, binned in the same way as the upper panel. Clearly, no strong evidence for deviations from $\daa=0$ remains after correcting the observations with the simple model of the long-range distortions.
        \label{fig:vlt-split-linear-results}}
    \end{figure*}

    After simulating the absorption systems used in \citet{2012MNRAS.422.3370K} using the information in their table A1, we applied the simple distortion model in \Fref{fig:vlt-split-linear-model} and fit for \daa as described in the first part of this Section.
    The code to run these simulations is freely available\footnote{Available at \href{https://github.com/jbwhit/AstroTools}{\url{http://github.com/jbwhit/AstroTools}}} and is flexible enough to allow anyone interested to run similar tests themselves.
    The upper panel of \Fref{fig:vlt-split-linear-results} shows the simulated values of \daa along with the binned VLT quasar measurements from \citeauthor{2012MNRAS.422.3370K} using their 13 redshift bins (chosen such that each bin contains a similar number of absorbers).
    We also plot the average weighted mean of the simulated \daa values within each bin using the inverse-squares of the 1-$\sigma$ uncertainties in the quasar measurements as weights.
    Note that statistical errors in the simulated \daa values themselves are negligibly small.

    The pattern of \daa values in \Fref{fig:vlt-split-linear-results} shows remarkable similarities to those of \citet{2012MNRAS.422.3370K}. 
    At redshifts $\zab<1.5$, where \ion{Mg}{ii} and \ion{Fe}{ii} transitions are most common, the positive slope of the simple model produces mostly negative \daa values, as seen in the \citeauthor{2012MNRAS.422.3370K} results. 
    Similarly, the general reversal in sign of \daa at higher redshifts is reproduced by the model, as well as the fact that the average magnitude of \daa is similar at the low and high redshift ends. 
    The simple model also predicts that \daa should change markedly $\zab\approx0.8$ where the \ion{Mg}{ii} transitions fall in the red arm while the \ion{Fe}{ii} transitions fall on the blue arm, i.e.~the narrow redshift range where the velocity shift between the \ion{Mg}{ii} and \ion{Fe}{ii} transitions reverses sign. 
    It is intriguing and perhaps telling that this matches the large deviation in the mean \daa in the observations at $\zab\approx0.8$. 
    On the other hand, the model does not reproduce the similar apparent positive deviation in \daa at $\zab\approx1.6$, nor does that deviation appear to be explained by the gap between the arms in other commonly used UVES wavelength settings we explored. 

    The lower panel of \Fref{fig:vlt-split-linear-results} shows the binned weighted mean results from the quasar spectra of \citet{2012MNRAS.422.3370K} after correcting each absorber's \daa value with its corresponding simulated value from the simple distortion model. The binning and weighting followed the same procedure as the upper panel\footnote{Based on the observed scatter around the weighted mean \daa, \citet{2012MNRAS.422.3370K} added $\sigma_{\rm rand}=9.05$\,ppm in quadrature to the statistical uncertainty of each absorber's \daa value. We recomputed $\sigma_{\rm rand}$ using the same methodology after correcting the \daa values with the simulated ones from our model. The new value used for the lower panel of \Fref{fig:vlt-split-linear-results} was $\sigma_{\rm rand}=8.58$\,ppm.}. No evidence for a deviation in \daa from zero remains after the correction. For example, the weighted mean \daa drops from $+2.1\pm1.2$\,ppm reported in \citeauthor{2012MNRAS.422.3370K} to $+0.8\pm1.2$\,ppm. This, of course, assumes that uncertainties in the model correction are negligible, which is unlikely to be true in detail. Nevertheless, two other factors provide some additional confidence that the model broadly reflects the distortions likely to be present in the VLT--UVES quasar spectra:
\begin{itemize}
\item The individual \daa values from the quasar spectra and simulated spectra correlate strongly. The 153 \daa measurement--simulation pairs have a Spearman rank correlation coefficient of 0.25 with an associated probability of $p=0.2$\,percent of by-chance correlation (i.e.~$\approx$3.1-$\sigma$ significance);
\item Correcting the quasar \daa values with those from the simple distortion model reduces the scatter. The root-mean-square (RMS) deviation from the mean \daa in the quasar sample is 27.4\,ppm; this reduces to 26.8\,ppm after subtracting the individual simulated \daa values from the corresponding quasar measurements. To estimate the significance of this reduction, we constructed 10,000 realizations of the simulated sample by selecting randomly, with replacement, 153 values from the simulated sample for each realization. This randomizes the association between absorbers and the corresponding corrections to \daa. Each realization of the simulated sample was used to correct the quasar \daa values and the RMS was computed as before. This reveals that a reduction to an RMS of $\le$26.8\,ppm occurrs only $p=0.3$\,percent of the time. That is, the reduction in the RMS we observe has a $\approx$3.0-$\sigma$ significance. Recognising that the mean \daa is redshift-dependent in the quasar results in \Fref{fig:vlt-split-linear-results}, we conducted the same test in three redshift bins. Similar reductions in the RMS were found, though with lower statistical significance: 20.5 to 19.8\,ppm with $p=2$\,percent for $\zab\le1.241$, 32.6 to 32.0\,ppm with $p=9$\,percent for $\zab\ge1.857$, and no reduction from 27.3\,ppm for the middle redshift bin with $p=37$\,percent.
\end{itemize}

    It is important to emphasize that we have adopted a single distortion slope to characterize our model of the distortions in the UVES quasar spectra, whereas we do find considerable variation in this parameter in the rather sparse supercalibration information from the data archive over the period before 2009. 
    Scaling the slope used in the model will simply scale the deviation of the model \daa points away from zero in \Fref{fig:vlt-split-linear-results}; changing its sign will reverse the deviations away from zero. 
    A more appropriate model would have to take into account information about the distortion slope for each of the many different exposures, sometimes in different wavelength settings, for each quasar in the \citet{2012MNRAS.422.3370K} study, but such a rich database of supercalibration exposures does not exist. 
    Nevertheless, even with the sparse information at hand, and the consequently simple model used, it appears that the long-range distortions have the characteristics to adequately explain the \daa UVES values measured in \citet{2012MNRAS.422.3370K}.

    In particular, our simple model does not incorporate any effect that should cause a correlation between the \daa value and the quasar sky position -- particularly declination -- that might explain the $\sim$2-$\sigma$ evidence, within the UVES results alone, for the dipole-like variation in \daa across the sky identified in \citet{2012MNRAS.422.3370K}. Indeed, after correcting the UVES quasar results with the simulated ones from our model, and following the same dipole analysis in \citeauthor{2012MNRAS.422.3370K}, we find very little change to the best-fit dipole$+$monopole model parameters, with the statistical significance of the dipolar variation (beyond a monopole-only model) changing only from $\approx$2.3 to $\approx$2.1-$\sigma$. However, the clear connection between the UVES derotator angle and changes in the ThAr lamp spectra seen in \Fref{fig:derotator-relationship} may ultimately prove important. That connection, combined with the fact that almost all the \citeauthor{2012MNRAS.422.3370K} quasar spectra were calibrated with subsequent, day-time ThAr spectra (i.e.~at a small number of standard, fixed derotator angles), leaves open the possibility of a subtle correlation between distortion slope and quasar declination in UVES. We have not yet identified any direct evidence for such a correlation.


    \subsection{Keck--HIRES Simulation} 
    \label{sub:implications_hires}
    The Keck--HIRES data used in \citet{Murphy:2003em,Murphy:2004kl} were all taken during the single CCD chip era.
    The single chip era is further divided into two time periods: before and after the image rotator was installed in 1996, as illustrated in \Fref{fig:hires-slope-history}.
    Of the 140 absorption systems in the Keck sample, 77 were observed during this pre-rotator era. 
    Given that we currently have no physical explanation or model for the origin of the long-range distortions in HIRES, this leaves no reliable way to extrapolate the results of these few supercalibrations to the many quasar spectra taken in the pre-rotator era.
    In addition, and in contrast to VLT--UVES, the Keck--HIRES long-range velocity distortions are not as easily modeled.
    We find roughly linear slopes of both positive and negative sign, as well as evidence of a bend in the distortion (as seen in \Fref{fig:hires-singlechip}). 
    Our search for a relationship between distortion sign and observational parameters such as seeing, telescope altitude/elevation, etc., did not yield any trend (as with UVES).

\begin{figure}
        \includegraphics[width=1.0\columnwidth]{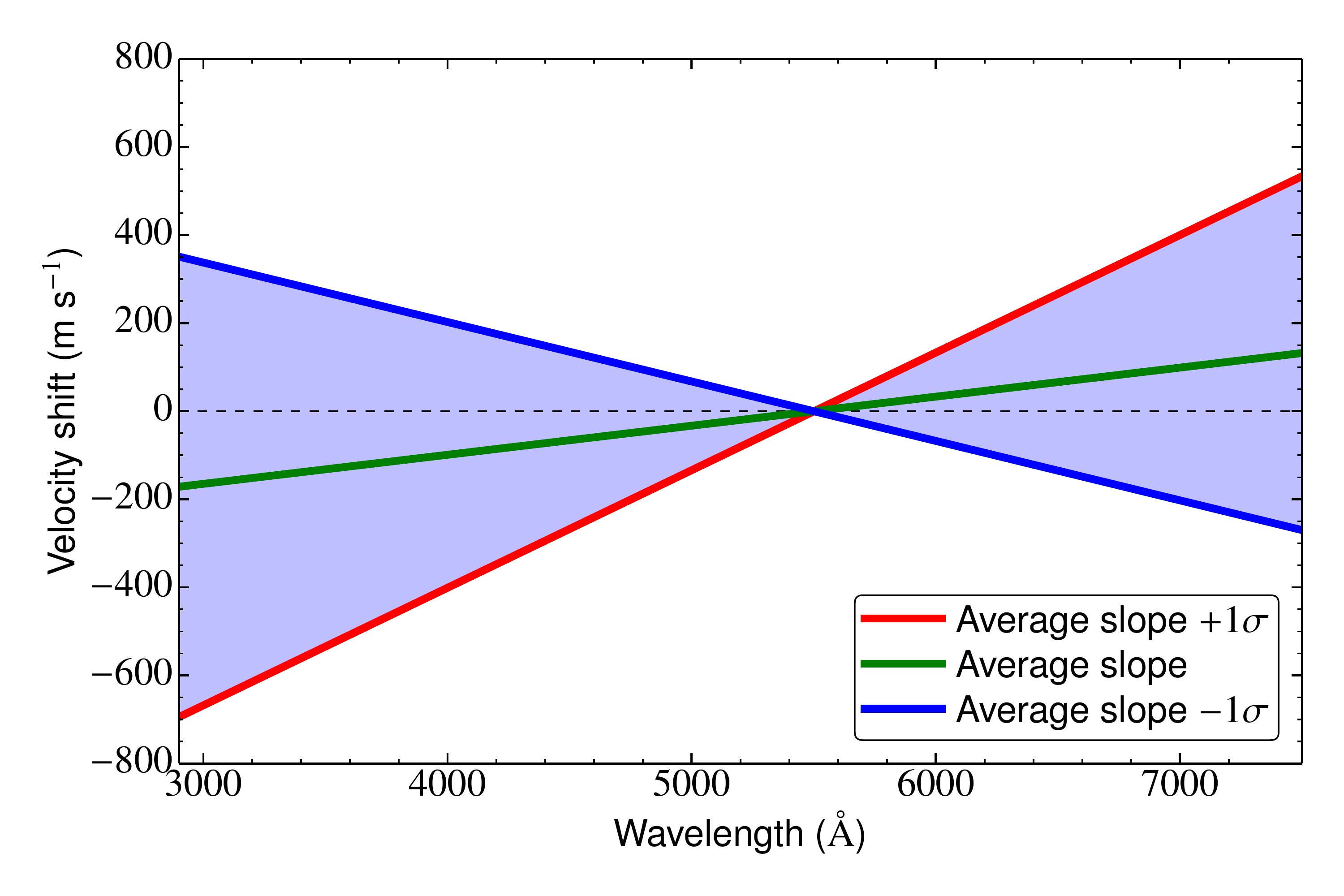}\vspace{-1.5em}
        \caption{The simple distortion functions used to model the HIRES velocity distortions.
        The three linear models have slopes of 66, 267 and $-135$\,\msperthousand.
        These values represent the average, $+$1-$\sigma$ and $-$1-$\sigma$ deviations in the nightly mean slopes observed in the 1-CCD chip era (see \Fref{fig:hires-slope-history}) in which the quasar spectra used by \citet{Murphy:2004kl} were observed.
        We apply this distortion model to the simulated HIRES data and plot the results in \Fref{fig:single-chip-single-night-linear}.
        \label{fig:hires-linear-model}} 
    \end{figure}

\begin{figure}
        \includegraphics[width=1.0\columnwidth]{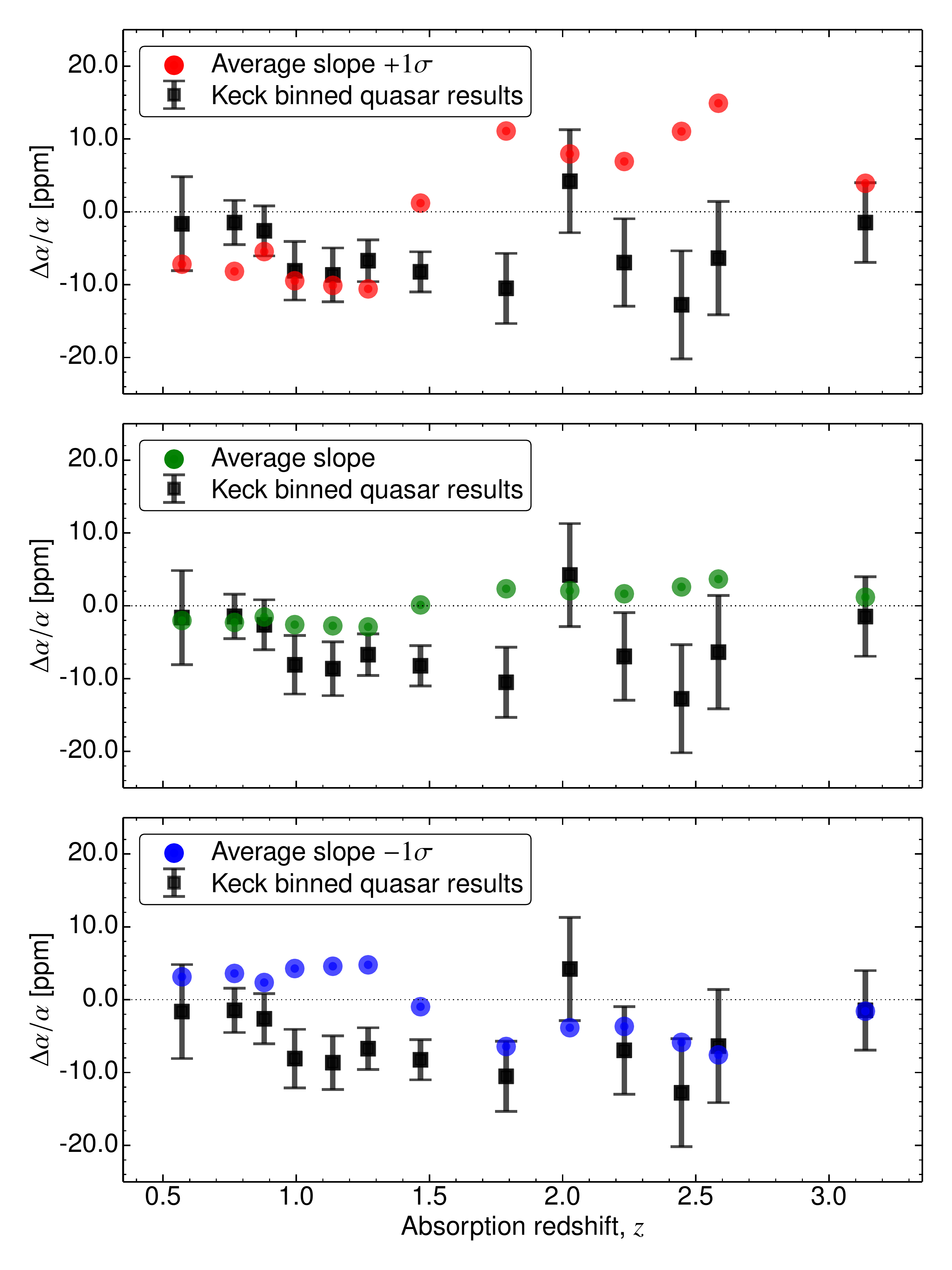}\vspace{-1.5em}
        \caption{
        Comparison between the original Keck--HIRES \daa measurements of \citet{Murphy:2003em,Murphy:2004kl} (black squares with 1-$\sigma$ errors) and the results of applying our simple distortion model (\Fref{fig:hires-linear-model}) to simulated spectra of those HIRES absorption systems (coloured/grey circles).
        Each bin contains $\approx$10 of the 140 \daa values from both datasets.
        The middle panel shows the results using a distortion model with a slope equal to the median nightly average of the long-range distortions observed prior to 2004, i.e.~66\,\msperthousand.
        The upper and lower panels show the results from models with the $+$1-$\sigma$ and $-$1-$\sigma$ slopes, respectively.
        No one model adequately explains all the Keck--HIRES \daa results at all redshifts, though the full range in distortion slopes is sufficient to explain the magnitude of the average deviations in \daa from zero.
        \label{fig:single-chip-single-night-linear}}
    \end{figure}

    We create the simple Keck model by characterizing the supercalibration distortions as simply as possible. 
    For each night of observation in the era where the data from \citet{Murphy:2004kl} was observed, we average the slopes for the supercalibration exposures, and take the median of these nightly averages.
    The average slope is 66 \msperthousand with a standard deviation of 201 \msperthousand. 
    We used this average value and the 1-$\sigma$ deviations as the slope for three simple HIRES models to apply to the simulated HIRES sample.
    The distortion models that we used are shown in \Fref{fig:hires-linear-model}.
    We apply these HIRES models to simulated absorption spectra corresponding to those used to measure \daa in \citet{Murphy:2004kl} \citep[i.e.~using the information in table A1 of][]{2012MNRAS.422.3370K}.
    Following the same procedure used for the UVES sample in \Sref{sub:implications_uves}, we plot the results in \Fref{fig:single-chip-single-night-linear}.

    The pattern of \daa values in \Fref{fig:single-chip-single-night-linear} shows that a linear model can reproduce \textit{either} the lower-redshift ($z<1.5$) region, or the higher-redshift ($z>1.5$) region, but not both at the same time. 
    A linear velocity distortion across the wavelength range corresponds to a simple compression or expansion of the wavelength scale, and \citet{Murphy:2003em} showed such models give opposite-signed behavior in \daa across the measured redshift range. 
    Our test here confirms that a single linear velocity distortion is not a good model for the Keck results and we do not find a single model that reliably reproduces the Keck measurements.
    On the other hand, it is important to note that, within the 1-$\sigma$ range of slopes, a linear distortion produces \daa values that encompass almost all of the measured results.
    Therefore, the long-range distortions observed in the Keck--HIRES supercalibration spectra have the general characteristics and magnitude to explain the possible evidence for a varying-$\alpha$ from HIRES quasar spectra. 
    However, the variability in the distortion slopes observed, combined with the uniformly negative \daa at all redshifts in the HIRES results, is the main barrier to these being a complete explanation at this stage.

    Another potential problem is the paucity of supercalibration information currently available for the pre-rotator era of HIRES (i.e.~before late 1996). Indeed, we only have supercalibration for two nights in that epoch, which accounts for over half of the Keck--HIRES absorber sample -- see \Fref{fig:hires-slope-history}. As discussed in \Sref{ssub:ssub:results_longrange_hires}, differential atmospheric refaction will have caused long-range distortions in the quasar and supercalibration spectra in the pre-rotator era. The possible effect of this on the Keck--HIRES \daa results was considered in detail by \citet{Murphy:2001uh,Murphy:2003em}. Despite simulations demonstrating that the effect should be detectable in the 77 affected absorbers, no correlation between \daa and zenith distance was observed. The pre- and post-rotator samples also gave very similar average \daa values. Thus, despite the possibility that pre-rotator HIRES spectra may have an additional, systematic contribution to their long-range distortions, evidence for it is not available from the very limited supercalibration information or the \daa results themselves.

    \subsubsection{High-contrast subsample} 
    \label{ssub:implications_hires_highcontrast}

    In the Keck measurements of \citet{Murphy:2004kl}, a subset of the \daa values were found to have a larger scatter than implied by their statistical uncertainties. 
    This subset of systems was at $\zab>1.8$ and there was significant additional scatter ``particularly around $\zab\approx1.9$''.
    This subset was found to correspond to absorbers in which a large number of transitions was used to determine \daa and, importantly, where there was a large range in the optical depths of these transitions.
    The large difference in the optical depths led to these systems being labeled ``high-contrast'' systems, with the remainder being labeled ``low-contrast''.
    An additional random error term of 20.9 ppm was added in quadrature to this ``high-contrast'' sample in \citet{Murphy:2004kl}.
    \citet{2012MNRAS.422.3370K} revisited and re-analyzed the same subsample. 
    Using a different method of `least trimmed squares', they found that the ``high-contrast'' sample required an additional 17.4 ppm  to be added in quadrature to the statistical errors for this subsample.

\begin{figure}
        \includegraphics[width=1.0\columnwidth]{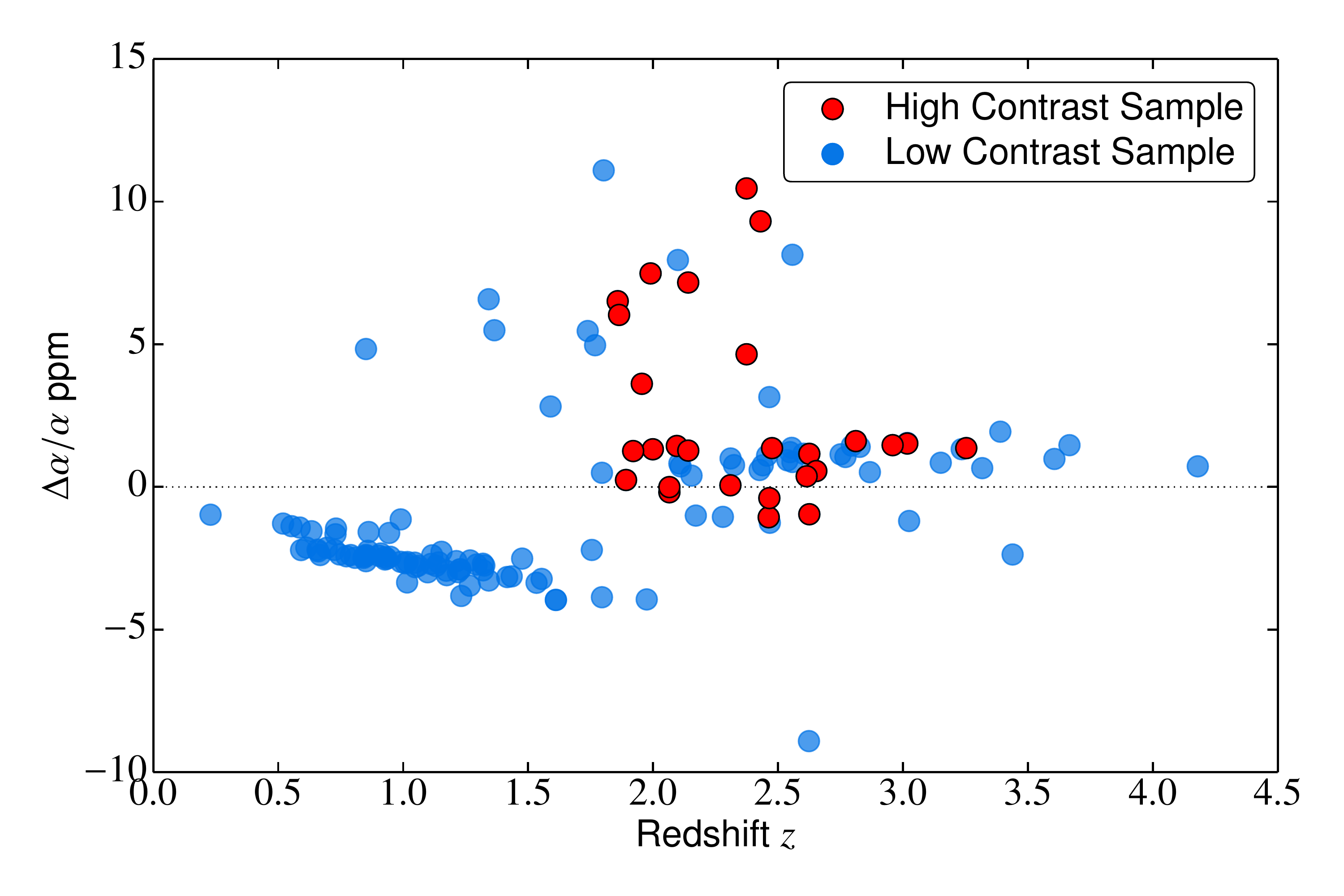}\vspace{-1.5em}
        \caption{The simulated HIRES \daa measurements of the average-slope distortion model (66\,\msperthousand).
        The red points are the 27 ``high-contrast'' points identified in \citet{Murphy:2003em,Murphy:2004kl}, and the remaining 113 are plotted in blue. \label{fig:high-contrast-sim}}
    \end{figure}

    \Fref{fig:high-contrast-sim} shows the effect of the average-slope model (66\,\msperthousand) velocity distortion applied to simulated data, with the red points highlighting the ``high-contrast'' sample.
    The simple linear model yields \daa results that increase in scatter for certain systems -- especially around the systems that lie at $\zab\approx1.9$.
    The extra scatter is clearly evident, and it suggests that even our simple models, constructed with a single velocity-component and only a linear distortion, create significant additional scatter around that redshift. 
    It is still possible that the explanation for the additional scatter described by \citet{Murphy:2003em,Murphy:2004kl} -- the difficulty in fitting a consistent velocity structure in such high-contrast absorbers -- explains the (remaining) additional scatter observed, but long-range distortions may be part of the explanation as well.

\section{Summary and conclusions} 
\label{sec:discussion}

We have presented a new method for testing the wavelength calibration
accuracy of high resolution spectrographs using exposures of `solar
twin' stars and an FTS solar reference spectrum. This
`supercalibration' method is similar to those based on iodine cell
stellar exposures and asteroid exposures presented previously and we
have improved those methods here as well. The solar twin and asteroid
supercalibration techniques provide a similar density of spectral
information as the iodine cell method but have the additional
advantage that the solar spectral features cover the entire wavelength
range of optical high resolution spectrographs
(e.g.~3000--10000\,\AA). The solar twin method presented here also
carries significant advantages over the asteroid method: solar twins are
located at fixed sky positions, at all right-ascensions, and many are
brighter ($\la9$\,mag) than most asteroids. It is therefore practical
and relatively efficient to observe a solar twin immediately after the
primary object of interest, in the same instrument set-up, with
minimal telescope slew, and with a very short exposure. In future,
this will enable more reliable checks on the standard ThAr wavelength
scale attached to the primary object exposures.

By applying the supercalibration technique to solar twin, asteroid and
iodine cell star exposures from both Keck--HIRES and VLT--UVES, we
find that significant distortions of the wavelength calibration are
ubiquitous in both spectrographs. The distortions are on scales of
both echelle orders ($\sim$50--100\,\AA, ``intra-order'') and entire
spectrograph arms ($\sim$1000--3000\,\AA, ``long-range''), and seem to
occur over the full variety of instrument setups and epochs. The
intra-order distortions have peak-to-peak amplitudes of up to 
$\sim$500\,\ms\ in both spectrographs. We find that the long-range
velocity distortions are approximately linear with wavelength in each
spectrograph arm. In UVES, they usually have a very similar linear
slope in each arm, with little velocity shift between the centres of
the two arms. In HIRES's single arm, many of the supercalibration
exposures show evidence for a somewhat non-linear distortion, typically with a
flattening in the bluest $\sim$3rd of the wavelength
range. Characterizing each spectrograph arm with a linear distortion,
we find similar magnitudes for the slope in both spectrographs:
typically $<200$\,\msperthousand but up to $\sim$800~\msperthousand
in some exposures. The sign of the slope seems to fluctuate in both
spectrographs over the 1--1.5 decades for which we have (sparse)
supercalibration information. The magnitude, and possibly the sign,
also fluctuates on very short (i.e.~intra-day) timescales. See
Figs.~\ref{fig:uves-slope-history} and \ref{fig:hires-slope-history}.

We explored the possible physical drivers of these distortions in both
spectrographs and established a working hypothesis for their cause:
differential vignetting of the astronomical object and ThAr
calibration light paths within the spectrograph, with the likelihood
that the ThAr vignetting dominates and is most variable. 
Some evidence supporting this hypothesis is that (i) the long-range and, at least
for UVES, intra-order distortions do not appear to change as the
position of the astronomical object within the spectrograph slit is
varied; (ii) the long-range slope for UVES changes significantly when
ThAr exposures taken at different (de)rotator angles are compared. If
this hypothesis is correct and the ultimate, mechanical cause can be
firmly established, it may be possible to accurately predict the
long-range distortions expected in previous quasar exposures. However,
at present, the available information is not sufficient to confirm our
hypothesis. The results of ongoing, dedicated tests on UVES and HIRES
will be reported in a future paper.

Finally, we estimated the approximate effect of the long-range
distortions on the measurements of \daa from large samples of archival
HIRES
\citep{Webb:1999cy,Webb:2001:091301,Murphy:2001uy,Murphy:2003em,Murphy:2004kl}
and UVES \citep{2011PhRvL.107s1101W,2012MNRAS.422.3370K} using
simulated quasar spectra. In the absence of a causal, physical model
to predict the long-range distortions, we instead modelled them with a
single, linear slope per spectrograph arm, as shown in
Figs.~\ref{fig:vlt-split-linear-model} and
\ref{fig:hires-linear-model} for UVES (2 arms) and HIRES (1 arm), respectively. The
slope assumed for each spectrograph was simply representative of the
significantly variable but sparsely sampled information about the
long-range distortions over the epochs in which the relevant quasar
spectra were recorded (Figs.~\ref{fig:uves-slope-history} and \ref{fig:hires-slope-history}). Even with these simplifying assumptions, the
effect on simulated quasar spectra mimics many of the features of the
UVES results obtained in \citet{2012MNRAS.422.3370K}, including a slow
change in the sign of \daa from $\zab\sim1$ to $\ga$1.8 and a sharp
change at $\zab\approx0.8$. Further, the simulated \daa values correlate strongly with the UVES quasar values, and correcting the latter with the former significantly reduces the scatter in \daa. However, the HIRES model does not
simultaneously reproduce a mean \daa of the same sign at both low and
high redshifts, as observed in the quasar results
\citep{Murphy:2003em,Murphy:2004kl}. Nevertheless, it may explain a
component of the additional scatter in the HIRES \daa measurements
obtained around $\zab\approx1.8$--2.4.

In summary, it is clear that long-range distortions of the wavelength scale identified here have the characteristics required to adequately explain the VLT--UVES quasar measurements of \daa, though they cannot entirely explain those from Keck--HIRES. That is, if both sets of quasar measurements are corrected with our distortion models, the VLT measurements are consistent with $\daa=0$ at all redshifts, while the Keck results deviate from zero at either $z<1.5$ or $z>1.5$ depending on the sign of the prevailing distortions. As a consequence, combining the datasets to search for subtle variations with redshift or across the sky, for example, requires a more detailed understanding of the possible systematic errors in the Keck--HIRES quasar measurements (at least). The long-range distortions therefore significantly weaken the possible evidence for variations in the fine-structure constant over cosmological time and spatial scales. Understanding the causes of the intra-order and long-range distortions, and whether they can be corrected, will also be crucial for future quasar measurements of \daa and for avoiding them in future spectrographs. In the interim, we recommend following each quasar exposure in such campaigns with both ThAr calibration and solar-twin supercalibration exposures to gauge the extent of, and allow correction for, long-range wavelength distortions in slit spectrographs.

\section*{Acknowledgments} 
\label{sec:acknowledgments}

    We gratefully acknowledge the referee, Paolo Molaro, for many insightful and helpful comments that improved this paper. We thank Juliet Datson and Christopher Flynn for discussions about solar twins, Julian King for assistance with dipole fitting, and
    Julija Bagdonaite, Neil Crighton, Tyler Evans, Glenn Kaprzack, Emily Petroff, David Lagattuta, Adrian Malec and Martin Wendt for numerous other helpful discussions.
    We also thank the Australian Research Council for \textit{Discovery Project} grant DP110100866 which supported this work.

    This research has made use of the Keck Observatory Archive (KOA), which is operated by the W. M. Keck Observatory and the NASA Exoplanet Science Institute (NExScI), under contract with the National Aeronautics and Space Administration.
    Some of the data presented herein were obtained at the W.M. Keck Observatory, which is operated as a scientific partnership among the California Institute of Technology, the University of California and the National Aeronautics and Space Administration. The Observatory was made possible by the generous financial support of the W.M. Keck Foundation.
    The authors wish to recognize and acknowledge the very significant cultural role and reverence that the summit of Mauna Kea has always had within the indigenous Hawaiian community.
    We are most fortunate to have the opportunity to conduct observations from this mountain.

    This research has also made use of the ESO UVES Paranal Observatory Project (ESO Program ID 266.D-5655) \citep{Bagnulo:2003vm}.
    The UVES and HARPS data presented herein were obtained from the ESO Science Archive Facility under request numbers (JWHITMORE): 
    105212, 105207, 105206, 102943, 102940, 102938, 102936, 88407, 88303, 88300, 88299, 88301, 88298, 81604, 81603, 81602, 14664, 174540, 174537, 11613, and 11578.

\bibliographystyle{mn2e}

\bspsmall

\label{lastpage}
\end{document}